

%
\magnification 1200
\vsize=7.5in
\hsize=5in
\voffset=0.35in
\hoffset=0.0in
\baselineskip 12pt plus 0pt minus 0pt

\newcount\eqnumber

\def\chapbegin#1{\centerline{\bf #1}\nobreak\bigskip
\eqnumber=1}

\def\sectbegin#1#2{\bigskip\bigbreak\leftline{\bf #1 #2}\nobreak
    \medskip\nobreak}

\def\nosectbegin#1{\bigskip\bigbreak\leftline{\bf #1}\nobreak\medskip}

\def\subhead#1{\medskip\goodbreak\noindent{\bf #1}\nobreak\vskip 0pt\nobreak}

\def\lapp{\hbox{$ {     \lower.40ex\hbox{$<$}
                   \atop \raise.20ex\hbox{$\sim$}
                   }     $}  }
\def\gapp{\hbox{$ {     \lower.40ex\hbox{$>$}
                   \atop \raise.20ex\hbox{$\sim$}
                   }     $}  }

\def\marbul{\strut\vadjust{\kern-2pt$\bullet$}}

\def\warning{}
\def\specialwarn{\vtop to
\strutdepth{\baselineskip\strutdepth\vss\llap{
\lower.1ex\hbox{$\bigtriangleup$}\kern-0.884em$\triangle$\kern-0.5667em{
!}\hskip 13.5pt}\null}}
\def\strutdepth{\dp\strutbox}


\def\new{{\the\eqnumber}\global\advance\eqnumber by 1}
\def\delaynew{{\the\eqnumber}}
\def\nownew{\global\advance\eqnumber by 1}
\def\last{\advance\eqnumber by -1 {\the\eqnumber}
    \global\advance\eqnumber by 1}
\def\eqnam#1{
\xdef#1{\the\eqnumber}}



\def\refto#1{~[{\it #1}]}
\def\refis#1{\item{#1}}
\def\references{}
\def\endreferences{}



\def\figure#1#2#3#4#5{
\topinsert
\null
\medskip
\vskip #2\relax
\null\hskip #3\relax
\special{illustration #1}
\medskip
{\baselineskip 10pt\noindent\narrower\rm\hbox{
Figure #4}:\quad
#5 \smallskip}
\endinsert}

\def\mfigure#1#2#3#4#5{
\midinsert
\null
\medskip
\vskip #2\relax
\null\hskip #3\relax
\special{illustration #1}
\medskip
{\baselineskip 10pt\noindent\narrower\rm\hbox{
Figure #4}:\quad
#5 \smallskip}
\endinsert}

\def\doublefigure#1#2#3#4#5#6#7{
\topinsert
\null
\medskip
\vskip #3\relax
\null\hskip #4\relax
\special{illustration #1}
\hskip #5
\special{illustration #2}
\medskip
{\baselineskip 10pt\noindent\narrower\rm\hbox{
Figure #6}:\quad
#7 \smallskip}
\endinsert}

\def\caption#1#2{
\baselineskip 10pt\noindent\narrower\rm\hbox{
#1}:\quad
#2 \smallskip}

\def\picture #1 by #2 (#3){
  \vbox to #2{
    \hrule width #1 height 0pt depth 0pt
    \vfill
    \special{picture #3} 
    }
  }

\def\scaledpicture #1 by #2 (#3 scaled #4){{
  \dimen0=#1 \dimen1=#2
  \divide\dimen0 by 1000 \multiply\dimen0 by #4
  \divide\dimen1 by 1000 \multiply\dimen1 by #4
  \picture \dimen0 by \dimen1 (#3 scaled #4)}
  }


\def\ma{m_{\rm a}}
\def\fa{f_{\rm a}}
\def\na{n_{\rm a}}


\chapbegin{Global String Radiation}
\vskip 10pt
\centerline{ R.$\,$A. Battye {~~and~~} E.$\,$P.$\,$S. Shellard}
\vskip 8pt
\centerline{ Department of Applied Mathematics}

\centerline{ and Theoretical Physics}

\centerline{ University of Cambridge}

\centerline{Silver Street, Cambridge epss$\,$@$\,$amtp.cam.ac.uk}\smallskip
\indent Paper
submitted to {Nuclear Physics B}.
\medskip
\medskip
\vskip 25pt
\centerline{ Abstract}
\medskip
{\narrower{
\baselineskip 10pt

\noindent We study Goldstone boson (or axion) radiation from
global  $U(1)$-strings making direct quantitative comparisons between numerical
scalar field
theory simulations and linearized analytic calculations in the dual
antisymmetric
tensor
representation.  Concentrating on periodic long string configurations, we
show excellent correspondence for both the amplitude and spectrum of radiation.
We also
find good agreement with a linearized analytic model for radiation
backreaction,
demonstrating that damping due to the lowest n=2 harmonic is effective in
strongly suppressing higher harmonics. This work establishes the
validity of the Kalb-Ramond action in describing low-energy global
string dynamics, along with the associated self-field renormalization of the
equations of motion. We also describe the different nature of massive radiation
from high curvature regions and the collapse and annihilation of a circular
loop.
As an
application in axion models, we discuss the relative loop and long string
radiation
contribution to a cosmological bound on the axion mass. A parallel linearized
formalism
for gravitational radiation is also provided. \smallskip}}  \smallskip
\baselineskip 12pt

\sectbegin{1~}{Introduction}

Global strings are associated with the breaking of a global symmetry that
leaves a degenerate vacuum manifold which is not simply-connected. Global
strings arise naturally in a variety of physical contexts; they are
generic in axion models and seemingly also in superstring theories, and
they can appear in GUT models, for example, through the breaking of a global
$B$$-$$L$
symmetry.  For appropriate mass scales, global strings can take on
many of the cosmological roles invoked for local gauge strings. Despite these
similarities, the literature on global strings is sparse and their physics is
inadequately understood and often even misunderstood. The chief distinction is
that, unlike local strings which radiate gravitationally, oscillating global
strings decay more rapidly into massless Goldstone bosons.  It is the nature
and characterization of this radiation which forms the subject of this paper.

The popularity of the axion in particle physics and cosmology has provided one
of the
main motivations for studying global strings, so it is enlightening to provide
a
context for this discussion.  The imposition of an extra global $U(1)_{PQ}$
symmetry
remains the most elegant solution to the strong $CP$-problem of QCD\refto{PQ}.
This Peccei-Quinn symmetry is broken
at some high energy scale $\fa$ and the corresponding pseudo-Goldstone boson
$\vartheta$
which appears is the axion\refto{Wil,Wei}. The axion field $\vartheta$ in these
extended
models becomes directly related to the previously arbitrary $\bar\theta$
-parameter of
QCD which is associated with any $CP$-violation. At the QCD-scale, ``soft''
instanton effects slightly tilt the symmetry breaking potential and the
axion acquires a small mass $\ma$.  The minimum of the potential then becomes
$\vartheta=0$ which corresponds to the $CP$-conserving value for $\bar\theta$.
Note
that neutron electric dipole measurements imply that $\bar\theta <
10^{-10}$.

In general, the couplings of the axion to ordinary matter are inversely
proportional
to the symmetry breaking scale $\fa$.  By making $\fa$ sufficiently large,
the axion
becomes effectively ``invisible.''
A lower bound for $\fa$ is provided in
astrophysical contexts by axion production because, being so weakly
interacting,
they
can escape from the entire volume of a star and compete with surface photon or
neutrino losses.  A study of red giant cooling provides the limit $\fa > 10
^9\,$GeV\refto{RG}, while supernova 1987a implies $\fa\gapp 10^{10}\,
$GeV\refto{SN}.

As one of the prime cold dark matter candidates, the cosmology of the axion has
attracted considerable interest. Requiring that the density contribution of
axions is
below critical (that is, $\Omega_{\rm a}<1$) provides an upper bound on $\fa$.
If the universe passes normally through the Peccei-Quinn phase
transition, then there are a
number of possible sources of axions.  Thermal axions produced by the phase
transition
are massless and are redshifted away by the expansion.  Another source is the
production of axions through oscillations about the minimum of the potential
$\vartheta
=0$ due to initial homogeneous misalignments of the axion above $T>
\Lambda_{\rm
QCD}$.  These zero-momentum axions were estimated to give the relative energy
density,\refto{AS,PWW},  \eqnam{\homdensity}
$$\Omega_a\approx 0.9\times
10^{\pm0.5} \bigg{(}{f_{a}\over10^{12}GeV} \bigg{)}\;\bar\theta_i^2\,
\eqno(\new)$$
where $\bar \theta_i$ is the initial average misalignment angle.\footnote{*}
{Note that
this bound can be weakened if there is substantial entropy production at some
time between the QCD scale and nucleosynthesis, for example, by engineering the
out-of-equilibrium decay of a massive particle species (such as ref.
\refto{Lyth1}).} These
axions were mistakenly thought to be the only significant contribution to the
axion
density, imposing the upper bound  \eqnam{\hombound} $$f_a \lapp
10^{12}GeV\,.\eqno(\new)$$

This cosmological estimate ignored topological effects at the Peccei-Quinn
phase
transition\refto{VE}, that is, the inevitable production of global
$U_{PQ}(1)$-strings through the
Kibble mechanism.\footnote{$\dagger$}{We should point out here that it is
possible to
avoid axion strings altogether through inflation, though such models may have
unsatisfactory aspects.  If $\fa$ is pushed above a GUT scale reheat
temperature,
we
must by (\hombound) anthropically arrange a small misalignment angle $\bar
\theta_i$
in our region of the universe.  The alternative of a low reheat temperature
below
$10^{12}$GeV assumes baryogenesis at the electroweak scale and may be ruled out
in future
by constraining the possible shape of the inflaton potential
using microwave anisotropies, given that one knows the relative contribution of
gravity waves
\refto{CKLL}.\smallskip} Davis\refto{Dava,Davb} suggested that the radiative
decay of the global string network provided a more significant contribution to
the
axion density.  This was estimated by assuming that direct radiation by long
strings into the fundamental harmonic was the primary energy loss mechanism.
Long
string radiation was subsequently studied in more detail by Davis \&
Shellard\refto{DSa} who demonstrated that  strings radiate with a classical
spectrum, predominantly in the second harmonic. Applying this spectrum with
the previous assumptions and numerical results from  network evolution, they
deduced a cosmological bound \eqnam{\stringbound} $$f_a \lapp
10^{10}GeV\,,\eqno(\new)$$
which is two orders of magnitude stronger than the zero-momentum bound
from (\homdensity).  A comparison with the astrophysical constraint brought
into
question the viability of axion models in such scenarios, suggesting that
inflation
may be required to eliminate axion strings.

There are, however, further contributions to the axion density.
When the axion mass switches on at $T = \Lambda_{\rm QCD}$, the $S^1$
vacuum manifold
is tilted, and the minima lie at the discrete values $\vartheta =0,
\pm 2\pi, \pm
4\pi, ...$   Thick domain walls form in regions where there are large
variations in the axion field $\vartheta$\refto{Sika}. In particular,
strings become
attached to domain walls which begin to dominate the dynamics of the string
network\refto{VE}.  Reconnections between strings and domain walls
facilitate the
rapid decay and annihilation of the hybrid system into more axions,
as demonstrated
numerically\refto{Sa,PRS,Sb}.  These additional axions will serve to
strengthen the
bound (\stringbound).   Paradoxically, axions produced by loops will
tend to weaken
the bound (\stringbound), which was derived assuming only long
string radiation.  If axions come predominantly from small loops instead of
long strings, they will be produced at higher frequencies and so lose more of
their relativistic energy through redshifting\refto{Sb}.  We shall return
briefly to this issue in Appendix A, leaving detailed analysis for a future
publication.

The orthodox classical understanding of string radiation assumed above has been
challenged by Sikivie and coworkers who have raised a number of important
issues\refto{HarS,HagS,Sikb}.  As we shall point out in \S2, the global string
appears to be a non-local object with logarithmic contributions to its energy
density coming from the Goldstone boson field on all length scales above the
string core width $\delta$.  Implicit in radiation calculations, therefore, is
the assumption that the global string maintains its integrity during its
motion, and that these less localized components merely renormalize the string
tension (provided that they are on length scales below the string curvature
radius).  Harari \& Sikivie\refto{HarS} questioned this renormalization and
suggested that the string core during its motion somehow decouples from the
surrounding ``cloud'' of Goldstone bosons.  They proposed that the global
string was in fact critically damped and lost all its oscillation energy in
less than one period, implying that the radiation power spectrum was flat with
contributions  at all wavelengths above $\delta$.  These suggestions were
supported by later numerical simulations of perfectly circular loops which, by
virtue of their symmetry, are in fact expected to behave in this
manner\refto{HagS}. If verified, a flat radiation spectrum would lead  to a
cosmological constraint, $\fa \lapp 10^{12}$GeV which is comparable to the
zero-momentum bound (\hombound) but at variance with original
string bound (\stringbound).

Here, we endeavour to settle this controversy beyond reasonable doubt; we make
an extensive study employing sophisticated numerical techniques which we
compare quantitatively with the analytic formalism developed to describe global
strings.  This formalism using two-index antisymmetric tensors
\refto{Wit,VVa,DSb}
has been used to study radiation from loop trajectories\refto{VVa,DQ}
and periodic
long strings\refto{Saka}. We point out where our analysis differs from previous
work and  we introduce  a simple backreaction model, based on the analytic
methods, as a first step in describing the damping of small-scale structure on
cosmic strings.   Resolution of these questions is all the more pertinent
because of ongoing and proposed axion dark matter searches.  The motivation for
these expensive experiments relies heavily on the theoretically predicted
parameter range allowed by the cosmological bounds (\hombound) and
(\stringbound).

Throughout this paper we employ a (+$\,---$) signature for the spacetime metric
$g_{\mu\nu}$, (+$\,-$) for the induced metric on the string worldsheet
$\gamma_{ab}$,
the coordinates for which are given by $X^{\mu} =
X^{\mu}(\sigma ,\tau)$, with the null coordinates,
$u=\sigma-\tau, ~v=\sigma+\tau$.

\sectbegin{2~}{Analytic formalism}

\subhead{The Goldstone model}
\smallskip
The essential features of global strings are exhibited in the simple $U(1)$
Goldstone model with action given by,
\eqnam{\actionone}
$$\eqalign{S &= \int\nolimits d^4x \bigg{\{}
\partial_{\mu}\bar\Phi\partial^{\mu}\Phi
 - {1\over4}\lambda(\bar\Phi\Phi - \fa^2 )^2 \bigg{\}}\cr
&= \int\nolimits d^4x \bigg{\{}(\partial_{\mu}\phi)^2 + \phi^2
(\partial_\mu\vartheta)^2 -
{1\over4}\lambda(\bar\Phi\Phi - \fa^2 )^2 \bigg{\}}\,,\cr} \eqno(\new)$$
where $\Phi$
is a complex scalar field which has been split as $\Phi = \phi e^{i\vartheta}$
into a
massive (real) Higgs component $\phi$ and a massless (real periodic) Goldstone
boson $\vartheta$.   The corresponding  Euler-Lagrange equations are given by
\eqnam{\elone} $$\partial_{\mu}\partial^{\mu}\Phi +
{1\over2}\lambda\Phi(\bar\Phi\Phi - \fa^2)  = 0\,. \eqno(\new)$$  For a
straight global string lying along the $z$-axis, we take the ansatz
\eqnam{\ansatz}
$$ \Phi(r, \theta) = \phi(r) e^{in\theta}\,,\eqno(\new)$$
where
$\theta$ is the azimuthal angle and $n$ is the winding number.
We take the usual
boundary conditions, $\phi(0) =0$ and $\phi\rightarrow\fa $ as
$ r\rightarrow\infty$.
Despite these conditions, the energy per unit length $\mu$ is mildly divergent
($n$$=$$1$), \eqnam{\stringxsect}
$$\mu(\Delta) ~\approx~ \mu_0 +  \int_\delta^\Delta \left [ {1\over r}{\partial
\Phi\over \partial \theta}\right]^2 2\pi r\, dr ~\approx~\mu_0 +
2\pi\fa^2 \ln \left
(\Delta\over \delta\right)\,,\eqno(\new)$$
where $\delta \sim (\sqrt \lambda \,\fa)^{-1}$ is the string core width and
$\mu_0\sim \fa^2$ is the core energy associated with the massive field
$\phi$ (that is, within $r \lapp \delta$).  The length-scale $\Delta$ is the
renormalisation scale provided in general by the curvature radius of the string
or the average inter-string separation. Typically, at the present day with $\fa
\sim 10^{12}$GeV the logarithm in (\stringxsect) is $\ln(\Delta/\delta) \approx
70$.  This implies that there is much more energy in the  Goldstone
field than in the string core $\mu_0$, a fact which has made the understanding
of global strings intuitively hazardous.

\subhead{Dual representation: antisymmetric tensors}
\smallskip
The analytic treatment of global string dynamics is hampered by the
topological coupling of the string to the Goldstone boson.  However, we can
exploit
the  well-known duality between a massless scalar field and a two-index
antisymmetric tensor field $B_{\mu\nu}$ to replace the Goldstone boson
$\vartheta$ in (\actionone) via the relation,
\eqnam{\duality}
$$\phi^2\partial_{\mu}\vartheta = {1 \over 2}f_a\epsilon_{\mu\nu\lambda\rho}
\partial
^{\nu}B^{\lambda\rho}\,. \eqno(\new)$$
The transformation in the presence of vortices where $\phi\rightarrow0$ has
to be
performed carefully\refto{DSb}, (\actionone) then becomes
\eqnam{\prekraction}
$$ S =
\int\nolimits d^4x\bigg{\{}(\partial_{\mu}\phi)^2
-{f_a^2 \over 6\phi^2}H^2 - {1 \over 4}\lambda(\phi^2 - f_a^2)^2\bigg{\}}
- 2\pi f_a\int\nolimits B_{\mu\nu}d\sigma^{\mu\nu}\,,\eqno(\new)$$
where the field strength tensor is $H^{\mu\nu\lambda} = \partial^{\mu}B^
{\nu\lambda} + \partial^{\lambda}B^{\mu\nu} +
\partial^{\nu}B^{\lambda\mu}$ and the area element $d\sigma^{\mu\nu}$ is
given in terms of the
worldsheet $X(\sigma,\tau)$ swept out by the zeroes of the Higgs field
($\phi=0$),
\eqnam{\areaelement}
$$d\sigma^{\mu\nu} = \epsilon^{ab}
\partial_aX^{\mu}\partial_bX^{\nu}d\sigma d\tau\,.\eqno(\new)$$
Effectively, the string acts as a source term for $B^{\mu\nu}$ with a
current density
given by  \eqnam{\current}
$$J^{\mu\nu} = {\fa\over 2}\int\delta^{(4)}[x-X(\sigma,\tau)]\,
d\sigma^{\mu\nu}\,.\eqno(\new)$$
 Integrating radially over the massive degrees of
freedom in (\prekraction) yields the Kalb-Ramond action\refto{KR},
\eqnam{\kraction}
$$S  = -\mu_0\int\nolimits \sqrt{-\gamma} \, d\sigma d\tau - {1 \over
6}\int\nolimits  d^4x\, H^2 - 2\pi f_a\int\nolimits
B_{\mu\nu}d\sigma^{\mu\nu}\,.\eqno(\new)$$ The first term is the familiar
Nambu action for local strings, the second term is the antisymmetric tensor
field strength for both external fields and the self-field of the string, and
the last term is a contact interaction between the $B^{\mu\nu}$ field and the
string worldsheet.  The coupling between the string and $B_{\mu\nu}$ is
analogous to the electromagnetic coupling of a charged particle, and is
amenable to the same calculational techniques.

Varying the action (\kraction) with respect to $X^{\mu}$ and $B_{\mu\nu}$
gives the equations of motion for the string and the antisymmetric tensor
field,  \eqnam{\Xeom}
$$ \eqalignno{&~~~~~~\mu_0(\ddot X^{\mu} - X^{\prime\prime\mu}) = 2\pi
f_aH^{\mu\alpha\beta} (\dot X_{\alpha}
X^{\prime}_{\beta} - X^{\prime}_{\alpha}\dot
X_{\beta})\,,&(\new)\cr
\eqnam{\Beom}
&\partial^{\alpha}\partial_{\alpha}B_{\mu\nu} = 2\pi f_a
\int\nolimits d\sigma d\tau(\dot X_{\mu} X^{\prime}_{\nu} -
X^{\prime}_{\mu}\dot X_{\nu})
\delta^{(4)}\big{(}x-X(\sigma,\tau) \big{)}\,,
&(\new)\cr}$$
where we have employed the conformal string gauge and the Lorentz antisymmetric
tensor gauge,
\eqnam{\conformal}
$$\eqalign{\dot X^2 +  X^{\prime 2} &=0\,,\cr
\dot X.X^{\prime} &=0\,, \cr
\partial _\mu B^{\mu\nu} &=0\,.}\eqno(\new) $$
As for the electron we have to perform a renormalization because the
antisymmetric
tensor field  comprises both the self-field  and the radiation
field of the string. The self-field is logarithmically divergent
but it can be absorbed such that\refto{LR,DQ}
\eqnam{\renormeom}
$$ \mu(\Delta)\big{(}\ddot X^{\mu} - X^{\prime\prime\mu}) =
f^{\mu}\eqno(\new) $$
where $\Delta$ is the renormalisation scale introduced in (\stringxsect)
and $f^{\mu}$
is the  finite radiation backreaction force on the string.
The exact form and effect of the radiation backreaction force term is
the subject of another publication\refto{BSb}.

The renormalised equations of motion $(\renormeom)$ for the string,
assuming the effects
of radiation backreaction to be small, that is $f^{\mu}\simeq 0$ can be
approximated by the Nambu equations of motion.
Using the conformal gauge (\conformal), these equations of motion have
the general solution  \eqnam{\nambueom}
$$ X^0 = t = \tau , \quad{\bf X} = {\textstyle{1 \over 2}}\big{[}{\bf a}(u) +
{\bf b}(v)\big{]}\,,\eqno(\new)$$
\noindent where $u$ and $v$ are null coordinates on the worldsheet and
\eqnam {\abdef}
$$ {\bf a}^{\prime 2} = 1,\quad {\bf b}^{\prime 2} =1\,. \eqno(\new)$$
\noindent These equations have closed loop solutions and long (or infinite)
string solutions. Fig.~1 illustrates two different long string solutions
which are periodic and will be employed in our subsequent analysis.

\figure{figure1.eps}{3in}{0.2in}{1}{Some specific periodic long string
solutions: (a)
helix, (b) pure sine wave}

\subhead{Radiation calculations using antisymmetric tensors}
\smallskip

Methods developed for calculating gravitational radiation from
strings\refto{Tur,VVb} \refto{Bur} are readily applied to an antisymmetric
tensor field
radiation\refto{VVa}.  This work focussed on loop radiation, but
the formalism has also been adapted for the study of long strings, that is,
infinitely long periodic sources\refto{Saka,Sakb}.  Given a periodicity
length $L$
in both $\sigma$ and $\tau$, then along a $z$-directed string we can
write $X^3 =
\alpha\sigma + P_1(u) + P_2(v)$, where the functions $P_1$ and $P_2$
are also periodic
with period $L$ and assumed to be small relative to $\alpha$.
By analogy with the gravitational radiation analysis\refto{Saka}, one can
deduce
that the average power radiated per unit length by the string is
\eqnam{\goldporig}
$${dP \over dz} = 2\pi\sum_{n=1}^{\infty}\omega_n\sum_{|\kappa_m|
<\alpha\omega_n}
\int_0^{2\pi}d\varphi\,\widetilde{J}^{\mu\nu*}(\omega_n,{\bf
k}^{\perp},\kappa_m) \,\widetilde{J}_{\mu\nu}(\omega_n,
{\bf k}^{\perp},\kappa_m)\,,\eqno(\new)$$
where $\omega_n = {2\pi n/ L}$, $\kappa_m = {2\pi m/\alpha L}$, ${\bf
k}^{\perp} = |{\bf k}^{\perp}|(\cos\varphi ,\sin\varphi)$, $|{\bf k}^{\perp}|$
=
$\sqrt{\omega_n^2 - \kappa_m^2}$ and $\widetilde J^{\mu\nu}$ is the Fourier
transform of
the source distribution $J^{\mu\nu}$ given by (\current). Using the split
into left- and
right-movers for $J^{\mu\nu}$ derived in Apppendix B one can deduce that
\eqnam{\goldplr}
$${dP \over dz} = {8\pi^2f_a^2 \over \alpha^2L}\sum_{n=1}^{\infty}n\sum_{|m|<
\alpha n \atop  m+n\,{\rm even}} \int_0^{2\pi} d\varphi \big{\{}|U|^2|V|^2 -
|U^*.V|^2\big{\}}\,,
\eqno(\new)$$
where
\eqnam{\defufull}
$$\eqalign{U^{\mu} &= \int_0^L{du \over L} \partial_u
X_R^{\mu}e^{-ik.X_R(u)}\,,\cr
V^{\mu} &= \int_0^L{dv \over L} \partial_v
X_L^{\mu}e^{-ik.X_L(v)}\,\cr
k^{\mu} &= (\omega_n,{\bf
k}^{\perp},\kappa_m)\,.}\eqno(\new)$$
As described in the previous section, the antisymmetric tensor
notation in principle allows the calculation of the radiation power for
particular
solutions of the Nambu equations of motion. One important  consequence of
this formula
is the fact that there is no contribution to the radiation power from the
fundamental
mode ($n$$=$$1$). (Note that the corresponding expressions for gravitational
radiation power are given in Appendix C.)

\subhead{Linearized radiation power expression}
\smallskip
Unfortunately, it is still difficult in most cases
to calculate the radiation power using (\goldplr). However, this expression
can be
linearized by
considering only first order terms in the modulus of radial
perturbations\refto{Hina}. One infinite periodic solution of (\nambueom)
and (\abdef) is
\eqnam{\linearX}
$$X^{\mu} = (\tau,{\bf X}^{\perp}, \alpha\sigma + P_1(u) + P_2(v))\,.
\eqno(\new)$$
In linearizing, we shall ignore powers of $|X^{\perp}|^2$ or greater.
The actual
radiation expressions previously derived for gravitational radiation are
not correct.
Here we deduce the analogous expression for Goldstone boson radiation from
a global
string, while the corrected result for gravitational radiation can be
found in Appendix C.

The gauge constraints (\conformal), in null worldsheet coordinates $(u,v)$,
can be written
as
\eqnam{\gaugecon}
$$ (\partial_u X^3)^2 + (\partial_u {\bf X}^{\perp})^2 =
{\textstyle{1 \over 4}},
\quad(\partial_v X^3)^2 + (\partial_v {\bf X}^{\perp})^2 =
{\textstyle{1 \over 4}}\,.\eqno(\new)$$
Rearranging and using a Taylor series expansion gives
\eqnam{\approxX}
$$ \partial_u X^3 \simeq {\textstyle{1 \over 2}}\big{(} 1 - 2(\partial_u
{\bf X}^{\perp})^2
\big{)},\quad \partial_v X^3 \simeq {\textstyle{1 \over 2}}\big{(} 1 -
2(\partial_v {\bf X}^{\perp})^2
\big{)}\,.\eqno(\new)$$
Hence to first order, $X^3 = (u+v)/2$ so $\alpha=1$, $P_1=0$ and $P_2=0$. We
can  write the linearized expression for $X_R$ and $X_L$ as
\eqnam{\xrxl}
$$ X_R^{\mu}(u) = \big{(}-{\textstyle{1 \over 2}}u,{\bf X}_R^{\perp}(u),
{\textstyle{1 \over 2}}u \big{)}\quad
X_L^{\mu}(v) = \big{(}{\textstyle{1 \over 2}}v,{\bf X}_L^{\perp}(v),
{\textstyle{1 \over 2}}v\big{)}\,.\eqno(\new) $$
These expressions can be inserted into (\defufull)  for $U^{\mu}$
and  $V^{\mu}$. Integrating $U^0$, $V^0$, $U^3$ and $V^3$ by parts and
linearizing
the exponential then gives
\eqnam{\uapprox}
$$\eqalign{U^{\mu} &= \bigg{(}{L \over 2\pi (m+n)} {\bf k}^{\perp} .
{\bf U}^{\perp}
,  {\bf U}^{\perp} , {-L \over 2\pi (m+n)} {\bf k}^{\perp} .
{\bf U}^{\perp} \bigg{)}
\,,\cr
V^{\mu} &= \bigg{(}{L \over 2\pi (n-m)} {\bf k}^{\perp} . {\bf V}^{\perp} ,
{\bf V}^{\perp} , {L \over 2\pi (n-m)} {\bf k}^{\perp} . {\bf V}^{\perp}
\bigg{)}
\,,}\eqno(\new)$$
where
\eqnam{\defuperp}
$$\eqalign{{\bf U}^{\perp} &= \int_0^L {du \over L} \partial_u
{\bf X}_R^{\perp} e^{
i\pi(m+n)u/L}\,,\cr
{\bf V}^{\perp} &= \int_0^L {dv \over L} \partial_v {\bf X}_L^{\perp}
e^{ i\pi(m-n)v
/L}\,,\cr
{\bf k}^{\perp} &= | {\bf k}^{\perp}|{\bf \hat n} = | {\bf k}^{\perp}|
(\cos \varphi,
\sin \varphi)\,.}\eqno(\new)$$
Therefore, one can deduce expressions for $|U|^2$, $|V|^2$, $|U^*.V|$ and
$|U.V|$ in terms of the linearized variables ${\bf U}^{\perp}$ ,
${\bf V}^{\perp}$ and $
{\bf \hat n}$, that is,
\eqnam{\uvperpsq}
$$\eqalign{|U|^2 &= -|{\bf U}^{\perp}|^2,\quad |V|^2 = -|{\bf
V}^{\perp}|^2\,\cr
U^*.V &= 2({\bf \hat n.U^{\perp *}})({\bf \hat n.V^{\perp}}) - {\bf U^{\perp
*}.
 V^{\perp}}\,,\cr
U.V &= 2({\bf \hat n.U^{\perp}})({\bf \hat n.V^{\perp}}) - {\bf U^{\perp}.
 V^{\perp}}\,.}\eqno(\new)$$
We are now in a position to evaluate the $\varphi$-integration as the only
remaining dependence  is in ${\bf \hat n}$. The following formulae for
{\bf W,X,Y,Z}
(all two dimensional vectors) are easily obtained,
\eqnam{\xy}
$$\int_0^{2\pi}d\varphi\,({\bf \hat n.X})({\bf \hat n.Y}) = \pi{\bf
X.Y}\eqno(\new)$$
\eqnam{\wxyz}
$$\eqalign{\int_0^{2\pi}d\varphi\,({\bf \hat n.W})({\bf \hat n.X})
({\bf \hat n.Y})
({\bf \hat n.Z}) &=\cr {\pi\over 4}\bigg{(}&{\bf (W.X)(Y.Z) + (Z.W)(X.Y) +
(X.Z)(W.Y)}\bigg{)}\,.\cr}\eqno(\new)$$
Applying these formulae to the power expression (\goldplr) with (\uvperpsq)
yields \eqnam{\goldlinear}
$${dP \over dz} = {8\pi^3f_a^2 \over L}\sum_{n=1}^{\infty}n\sum_{|m|<
 n \atop  m+n\, {\rm even}} \big{\{}|{\bf U}^{\perp}|^2|{\bf V}^{\perp}|^2 +
|{\bf U}^{\perp *}.{\bf V}^{\perp}|^2 - |{\bf U}^{\perp}.{\bf V}^{\perp}|^2
\big{\}}\,.\eqno(\new)$$
If one of the components
of ${\bf X}^{\perp}$ is zero then the linearized power expression can be
simplified
further \eqnam{\goldlinspec}
$${dP \over dz} = {8\pi^3f_a^2 \over L}\sum_{n=1}^{\infty}n\sum_{|m|<
 n \atop  m+n\, {\rm even}}|{\bf U}^{\perp}|^2|{\bf V}^{\perp}|^2\,.
\eqno(\new)$$
The linearized power expression $(\goldlinear)$ will only give the leading
order term if the
average value of $|{\bf X}^{\perp}|$ is sufficently small, note also that
it is possible that there is no contribution at this order.  Note also from
(\goldplr) or (\goldlinear) that left-moving modes propagating along a
straight string will not radiate if right-moving modes are absent (and vice
versa).

\subhead{Radiation power for specific solutions}
\smallskip
Long string configurations can be quantified by the wavelength,
$\lambda$ $(=L)$ and the  amplitude, $A$. For solutions of the form
(\linearX), the important parameters are the ratio of the amplitude to
wavelength, $\varepsilon=2\pi A/\lambda$ (denoted the relative amplitude),  and
the oscillation frequency, $\Omega=2\pi/\lambda$.
In particular, we note that solutions for which $\varepsilon>1$ are not
allowed by  (\abdef) in our linearized description.  Of course,
solutions with $\varepsilon>1$ are allowed for more general string
configurations, but in the cases of high symmetry below they
become highly  degenerate with whole sections of string approaching luminal
velocities (see the discussion in section 3).

\smallskip \noindent{\it (a)
Helix solution} \smallskip

\figure{figure2.eps}{3.3in}{0in}{2}{Log-linear power spectrum of the helix
solution
at small ($\varepsilon=0.2$) and large ($\varepsilon=0.99$) amplitude.}

A simple, symmetric solution of the form (\linearX) can be constructed from
equal and oppositely propagating sine waves,
\eqnam{\helixdef}
$$ {\bf X} = \bigg{(}{\varepsilon \over 2\Omega}\big{[}\cos\Omega u +
\cos\Omega v
\big{]},{\varepsilon \over 2\Omega}\big{[}\sin\Omega u + sin\Omega v \big{]},
{1 \over 2}\sqrt{1-\varepsilon^2}\,(u+v)\bigg{)}\,,\eqno(\new)$$
where $0<\varepsilon <1$ and $\varepsilon \to 1$ in the relativistic limit.
This
corresponds to a helicoidal solution which oscillates between a static helix
and a
straight line.

In this case we can use the full nonlinear analysis $(\goldplr)$. The
radiation power  expression contains many Bessel functions with argument
$\tilde k = {1 \over 2}k\sqrt {n^2 -{m^2 /(1 -\varepsilon^2)}}$,

\eqnam{\fullhelixpower}
$$ {dP \over dz} = \sum_{n=1}^{\infty}P_n = \sum_{n=1}^{\infty}
\sum_{ |m|<n\sqrt{1-\varepsilon^2}
\atop m+n \, {\rm even}} \tilde P_{n,m}\,,\eqno(\new)$$
where
\eqnam{\harmpower}
$$\eqalignno{\tilde P_{n,m} = &{\pi^2\Omega f_a^2 \over 1-\varepsilon^2}
\bigg{[}-
{\textstyle{1 \over 4}}\varepsilon^4
\big{(}(J_lJ_{u+1}-J_uJ_{l+1})^2 + (J_uJ_{l-1}-J_lJ_{u-1})^2 -2J^2_lJ^2_u
&(\new)\cr
&-{\textstyle{1 \over 2}}(J_{l+1}J_{u-1}-J_{l-1}J_{u+1})^2\big{)} +
\varepsilon^2J_lJ_u( 2J_lJ_u - J_{l+1}J_{u-1} -
J_{l-1}J_{u+1})\,\,\,\bigg{]}\,,&\cr}$$ and
\eqnam{\deful}
$$l=\textstyle{1\over 2}(m+n)\,,\, u=\textstyle{1\over 2}(m-n)\,.\eqno(\new)$$
(This expression was
previously obtained\refto{Sakb}, but without the shortcut provided by the
left- and
right-moving split of Appendix B.)

It can be easily shown that $\tilde P_{n,0} =0$ for all positive integer
$n$. From this
and the conditions $|m| <n\sqrt{1-\varepsilon^2}$ and $m+n$ even, one can
deduce that
the lowest even harmonic radiating is given by
\eqnam{\neven}
$$ n_{\rm even}= {\rm min}\big{\{} n : n > {2 \over \sqrt{1-\varepsilon^2}},
\hbox{$n$ an even integer}\big{\}}\eqno(\new)\,,$$
\noindent and the lowest odd harmonic is given by
\eqnam{\nodd}
$$ n_{\rm odd}= {\rm min}\big{\{} n : n > { 1 \over \sqrt{1-\varepsilon^2}},
\hbox{ $n$ an odd integer}\big{\}}\eqno(\new)$$
\noindent For all $\varepsilon$, $n_{\rm even}$ is greater
than two and, for $\varepsilon<\varepsilon_{\rm crit}=0.94$, $ n_{\rm odd}$
is three.  Therefore for $\varepsilon<\varepsilon_{\rm crit}$, the third
harmonic is the dominant contribution to the radiation power of the helix. In
this case, one can deduce a linearized power  expression by calculating the
power in the third harmonic to first order in $\varepsilon$, which gives
\eqnam{\helixlinpower}
$${dP \over dz} \simeq {4\pi^3f_a^2\varepsilon^{10} \over 125L}.\eqno(\new)$$
\noindent Applying the linearized power expression (\goldlinear) to this
solution gives zero, as it only takes into account terms of
order $\varepsilon^4$ and below.

For $\varepsilon<\varepsilon_{\rm crit}$, one can see from fig.~2 that the
power in the nth harmonic decays exponentially, with exponent proportional to
$n$  for large $n$ (that is, $P_n \propto e^{-\alpha n}$, $n>>1$). Fig.~3
illustrates how $\alpha$ varies with $\varepsilon$, showing that
the exponential fall-off $\alpha$ becomes larger at small $\varepsilon$. The
proposal of Sikvie {\it et al.\/} for radiation from long strings requires that
$P_n \propto n^{-1}$. This contention should also be contrasted with the
result for the Burden-Tassie loop solutions ($P_n \propto n^{-4/3}$) and kinky
loops ($P_n \propto n^{-2}$)\refto{VVa}.

The helix solution has a great deal of symmetry which probably causes
the cancellation of the second harmonic. Configurations of this type would
not be expected to form in a realistic string network, because oppositely
propagating modes will not have strong correlations. However, given that
the helix is so weakly radiating, it conceivably could be the endpoint
of a radiation process, say for small-scale structure on a large loop.  The
stability of this solution to radiation damping and other perturbations is,
therefore, an interesting issue.  Given that we do not observe the
cancellation of the second harmonic in more general solutions, we believe that
the helix is actually an exceptional result and will be unstable to stronger
radiative decay (a contention for which we provide numerical evidence later).
Nevertheless, it is not unreasonable to suggest that some helix properties,
particularly the exponential fall-off of the harmonics at large n, will
be generic for all $\varepsilon < 1$ long string solutions.

\figure{figure3.eps}{3in}{1in}{3}{$\alpha$ against $\varepsilon$  such that
$P_n
\propto e^{-\alpha n}$ for large n.}

\smallskip
\noindent{\it (b) Sine solution}
\smallskip

A solution, similar to the helix, but with perturbations in only one
direction is
\eqnam{\cosinedef}
$$ {\bf X} = \bigg{(}{\varepsilon \over 2\Omega}\big{[}\cos\Omega u +
\cos\Omega v
\big{]},0,{1 \over 2\Omega}\big{[}E(\varepsilon,\Omega u) +
E(\varepsilon,\Omega v)
\big{]}\bigg{)}\,,\eqno(\new)$$
\noindent where $E(k,\phi)$ is the incomplete elliptic integral of the
second kind,
defined by
\eqnam{\ellipicdef}
$$ E(k,\phi)=\int_0^{\phi}d\theta\sqrt{1-\varepsilon^2\sin ^2\theta}\,,
\eqno(\new)$$
where  $0<\varepsilon <1$ and $\varepsilon \to 1$ in the relativistic limit.

The elliptic function can be written as a linearly increasing function plus
a periodic function, so the solution can be seen to be in the form
(\linearX) and  we can employ the power expressions (\goldplr) or
(\goldlinear). Unfortunately, the full nonlinear version
(\goldplr) becomes too complicated, so we apply (\goldlinear) to obtain
\eqnam{\cosinelinpower}
$${dP \over dz} = {\pi^3f_a^2\varepsilon^4 \over 16L}\,.\eqno(\new)$$
\noindent This only yields the power up to order $\varepsilon^4$, for which
only contribution comes from the second harmonic ($n$$=$$2$). Presumably, as
for the helix, the power will become  dominated by higher order terms at large
$\varepsilon$, but it seems reasonable to assume that
there also exists some $\varepsilon_{\rm crit}$ below which $(\cosinelinpower)$
is a good approximation. In any case, for
small amplitude waves it is clear that the dominant contribution to the
radiation power comes from the second harmonic.
\smallskip \noindent{\it (c) Kink
solution} \smallskip

It is well known that (\abdef) has solutions with points at which the string
velocity $\dot {\bf x}$ and tangent ${\bf x}'$ are discontinuous; these
points, known as kinks, propagate along the string at the speed of light. Here
we give a simple solution for a periodic distribution of kinks on a long
string, consisting of left and right moving perturbations, $X_U$ and $X_V$
respectively,
\eqnam{\kinkdef}
$$ {\bf X} = \bigg{(}X_U+X_V,0,{1 \over 2}\sqrt{1-{4\varepsilon^2 \over
\pi^2}}(u+v)
\bigg{)}\,,\eqno(\new)$$
\noindent where
\eqnam {\kinkone}
$$X_U = \left\{\eqalign{
&{2\varepsilon  \over \pi}u \cr &{2\varepsilon  \over \pi}\bigg{(}{1 \over
2}L-u
\bigg{)} \cr
&{2\varepsilon  \over \pi}\bigg{(}-L+u\bigg{)}} \right.\hbox{  }
\eqalign{&0<u<{1 \over 4}L \cr &{1 \over 4}L<u<{3 \over 4}L \cr&{3 \over 4}
L<u<L\,,}
\eqno(\new)$$
\eqnam{\kintwo}
$$X_V = \left\{\eqalign{
&{2\varepsilon  \over \pi}v \cr &{2\varepsilon  \over \pi}\bigg{(}{1 \over
2}L-v
\bigg{)} \cr
&{2\varepsilon  \over \pi}\bigg{(}-L+v\bigg{)}} \right.\hbox{  }
\eqalign{&0<v<{1 \over 4}L \cr &{1 \over 4}L<v<{3 \over 4}L \cr&{3 \over 4}
L<v<L\,,}
\eqno(\new)$$
\noindent where $0<\varepsilon < {\pi \over 2}$ and, in this case,
$\varepsilon\to{\pi\over 2}$ is the relativistic limit.
\bigskip
\indent Once more the full non-linear radiation power expression
(\goldplr) is intractable, so we use the linearized
version (\goldlinear) which yields
\eqnam{\kinklinpower} $${dP \over dz} =
{8192\,f_a^2\varepsilon^4 \over \pi^5L}\sum_{n=1}^{\infty}n \sum_{|m|<n
\atop m+n\,{\rm even}}{1 \over (n^2-m^2)^2}\bigg{[}1 - \cos {\pi (m+n) \over
2}\bigg{]}\bigg{[}1 - \cos{\pi (m-n) \over 2}\bigg{]}\,.\eqno(\new)$$
\noindent Again, this expression should be valid for
$\varepsilon< \varepsilon_{\rm crit}$ for some $\varepsilon_{\rm crit}$, but
even for sufficently small  $\varepsilon$  the radiation power is a
logarithimically divergent  infinite sum (that is, $P_n \propto n^{-1}$ for
large $n$). This can be seen analytically from (\kinklinpower)
since the sum is dominated by the $m=n-2$ terms and numerically in fig.~4. The
kink solution $(\kinkdef)$ is an
infinite Fourier series over odd harmonics. However, in a physical context
this divergence is cut off initially at the string width $\delta$ and at
increasingly larger scales as radiation backreaction rapidly `rounds' the
kink.  This `rounding' will eliminate the more rapidly decaying high
frequency contributions, ultimately leaving only the fundamental mode
excitation.

\figure{figure4.eps}{3in}{0in}{4}{Log-log power spectrum for the kink
solution showing power law fall-off.}

\smallskip
\noindent{\it (d) Spectrum of sinusoidal solutions}
\smallskip

The following solution corresponds to a superposition of a large, but finite
number, of sinusoidal perturbations,
\eqnam{\sinedef}
$${\bf X} = \bigg{(}{\bf X}^{\perp},\sqrt{1-\varepsilon^2}\bigg{)}\,,
\eqno(\new)$$
\noindent such that
\eqnam{\sinedefone}
$${\bf X}^{\perp} = {1 \over 2}\sum_{N\, st \, N\not=0}{\varepsilon \over
N\Omega}
\bigg{[}{\bf A}_Ne^{iN\Omega u}+{\bf B}_Ne^{iN\Omega v}\bigg{]}\,,\eqno(\new)$$
\noindent where $A_N$ and $B_N$ satisfy
\eqnam{\constone}
$$\eqalign{&{\bf A}_N^*=-{\bf A}_{-N} ,\quad {\bf B}_N^*=-{\bf B}_{-N}\,,\cr
&\sum_{N\not =0}|{\bf A}_N|^2 = 1 ,\quad \sum_{N\not =0}|{\bf B}_N|^2 =
1\,,\cr &\sum_{N\not =0}{\bf A}_N.{\bf A}_{n-N} =0
,\quad\sum_{N\not =0}{\bf B}_N.{\bf B}_{n-N} =0\, \quad \forall
n\not=0\,.} \eqno(\new)$$
\noindent Such a random superposition more realistically models the
perturbations expected on a string network.
Using the linearized
radiation power expression (\goldlinear), we obtain
\eqnam{\sinelinpower}
$$\eqalign{{dP \over dz} =& {\pi^3 f_a^2\varepsilon^4\over 2L}\sum_{n=1}
^{\infty}n\sum_{|m|<n
\atop m+n\,{\rm even}}\bigg{[}|{\bf A}_{m+n \over 2}|^2|{\bf B}_{m-n \over
2}|^2
\cr &-|{\bf A}_{m+n \over 2}.{\bf B}_{m-n \over 2}|^2 +
|{\bf A}_{m+n \over 2}^*.{\bf
B}_{m-n \over 2}|^2\bigg{]}\,.}\eqno(\new)$$
\noindent This illustrates the generic $\varepsilon^4$ dependence of the
radiation power even for the higher oscillation frequencies.  In special
cases, such as the helix, the contribution at this order can cancel.

\subhead{Simple backreaction model}
\smallskip

The string configurations discussed in the previous section are solutions
of the Nambu equations of motion. In flat space they will continue to
oscillate forever at constant amplitude.  Even in an expanding universe, the
amplitude will only decay slightly due to Hubble damping once $\lambda <
H^{-1}$.  This is clearly physically unrealistic, as it takes no account of
the fact that such configurations radiate Goldstone bosons. Here, we will
endeavour to incorporate this radiation backreaction in some special cases.

In the previous section, we calculated the radiation power from several
solutions, giving special consideration to the contribution from higher
harmonics.  This was only significant at large relative amplitude
$\varepsilon$ or for `kinky' solutions.  Even in these situations, however,
our expectation is that these higher harmonics will be rapidly suppressed as
their amplitude decays or the kink is rounded.  Radiation power is soon
dominated by one particular mode.
For backreaction in this simple case, the expressions for
the radiation power include powers  of $\varepsilon$ and one can deduce a
phenomenological model for the decay of $\varepsilon$.

\eqnam{\brenergy}
Given that the
generic radiation power for any oscillation mode is proportional
to $\varepsilon^4$, we can model radiation damping by considering the
following equations,\footnote{*}{Note we implicitly assume that the cut-off
$\Delta$ for $\mu$ in (\brenergy) is comparable to the curvature radius of the
oscillating string.}
$$\eqalign{E &= \mu + \alpha\mu\varepsilon^2\,, \cr
{dP\over dz} &= -{dE \over dt} = {\beta\varepsilon^4 \over L}\,,}\eqno(\new)$$
where $E$ is the energy per  unit length and $\alpha$ is the
solution-dependent coefficent of $\varepsilon^2$ in the
power series expansion of
\eqnam{\energyunit}
$${1\over L}\int_0^L{dX^3\over d\sigma}d\sigma\,.\eqno(\new)$$
For the sinusoidal
solution $\alpha=\textstyle{1\over 4}$, whereas for the helix
$\alpha=\textstyle{1\over 2}$ (as well as $\beta=0$). In flat space, $\mu$ and
$L$ are constants so we can integrate the differential equation (\brenergy)
to obtain
\eqnam{\brnonsoln}
$$ \varepsilon = \varepsilon_0 \bigg{(}1 + {\beta\varepsilon_0^2 \over
\alpha\mu L}t\bigg{)}^{-1/2}
,\quad E = \mu + \alpha\mu\varepsilon_0^2\bigg{(}
1 + {\beta\varepsilon_0^2 \over \alpha\mu L}t\bigg{)}^{-1}\,.\eqno(\new)$$
where initially $\varepsilon (0) = \varepsilon_0$. Defining $A=\mu$,
$B=\alpha\mu\varepsilon_0^2$ and $C= {\textstyle {\alpha\mu L/
\beta\varepsilon_0^2}}$, this becomes\eqnam{\brnonsolnen}
$$ E = A + B \bigg{(}1 + {t \over C}\bigg{)}^{-1}.\eqno(\new)$$
Initially, the decay
is linear $(t<<C)$, while asymptotically at late times $(t>>C)$ the energy
decays as a power law, $t^{-1}$. Fig.~5  shows the oscillation energy against
time for some reasonable values of $A$, $B$ and $C$.
A useful property of the model is the effective
half-life of the oscillation energy,
\eqnam{\halflife}
$$t_{1/2}=\alpha\mu
L/\beta\varepsilon_0^2\,.\eqno(\new)$$
The half-life depends on the initial
relative amplitude $\varepsilon_0$, increasing rapidly as $\varepsilon_0$
decreases.

\figure{figure5.eps}{3in}{-0.1in}{5}{Oscillation energy against time For A=10,
B=100 (unbroken line) and B=1000 (broken line).}

The power law decay (\brnonsolnen) of the oscillation energy provides a useful
quantitative test of this formalism (see section 3), but it is a special case
depending upon the left- and right-moving modes having equal amplitude.
Strings in a realistic network will in general have a random superposition of
unequal left- and right-moving modes.  We can demonstrate heuristically that
exponential decay will be generic for such modes.   Suppose that
$\varepsilon_L$ and $\varepsilon_R$ are the `averaged'  amplitudes of the
left- and right-moving waves respectively.  Motivated by
(\goldlinspec) we might expect these modes to obey a cross-coupled
version of (\brenergy)\refto{Hinb}, that is,
\eqnam{\genenergy}
$$\eqalign{\alpha\mu{d\varepsilon_R^2\over dt} =
- {\beta\varepsilon_L^2\varepsilon_R^2\over L}\,,
 &\quad \varepsilon_R (0) = A\,,\cr
\alpha\mu{d\varepsilon_L^2\over dt} = -
{\beta\varepsilon_L^2\varepsilon_R^2\over L}\,,
 &\quad \varepsilon_L (0) = B\,.\cr} \eqno(\new)$$
The solution we find in this
case when $A\ne B$ is \eqnam{\gensoln}
$$\eqalign{& \varepsilon_R^2 = {A^2(B^2-A^2)\over B^2e^{\beta (B^2-A^2)t
/ \alpha\mu L} -A^2}\,,\cr & \varepsilon_L^2 = {B^2(B^2-A^2)\over B^2 -
A^2e^{-\beta (B^2-A^2)t / \alpha\mu L}}\,,\cr} \eqno(\new)$$
The asymptotic behaviour of these solutions is of interest. In particular,
we find that for $B>A$, $\varepsilon_R^2\to 0$ and $\varepsilon_L^2\to
B^2-A^2$ exponentially, whereas for $B<A$, $\varepsilon_L^2\to 0$ and
$\varepsilon_R^2\to A^2-B^2$. Whichever of the movers initially
has the largest amplitude will approach a finite amplitude, while the other
will decay to zero exponentially.  The oscillation energy decay
timescale again depends on the relative initial amplitudes.

\sectbegin{3~}{Numerical methods and results}

\subhead{Algorithms and boundary conditions}
\smallskip

As we have seen there is a duality between the Kalb-Ramond action (\kraction)
and
the Goldstone action
(\actionone) in the presence of vortices. This duality implies that
the two actions should exhibit the same dynamics. In the previous section the
dynamics
of various long string configurations were examined within the framework of the
Kalb-Ramond action.  Here we directly compare the results of simulations with
the
same configurations in the Goldstone model.  We tackle this problem numerically
because, as we have noted, the Goldstone action
does not lend itself to simple analytic
calculations due to the topological coupling of the massless field.

We have developed sophisticated numerical algorithms to allow the simulation
of long string configurations within the framework of the Goldstone model. As
in
ref.\refto{Sc}, we  discretize space on a three-dimensional grid with
dimensions
$N_1,N_2, N_3$ in the $x,\,y,\,z$ directions respectively, solving the rescaled
($f_a\to
1,~ \lambda\to 2$) Euler-Lagrange equation,
\eqnam{\elrescale}
$${\partial^2\Phi\over\partial t^2} - {\partial^2\Phi\over\partial x^2} -
{\partial^2\Phi\over\partial y^2} - {\partial^2\Phi\over\partial z^2}
+\Phi(\bar\Phi \Phi-1) = 0\,.\eqno(\new)$$
We employ a second-order leapfrog algorithm for the time derivative
and second- or
fourth-order finite difference approximations for the spatial derivatives
(generally the latter).  We denote
the two algorithms as 2--2 and 2--4 respectively; they are discussed in greater
detail elsewhere\refto{BSa} (see also ref.\refto{Sd}).

The cylindrically symmetric straight string ansatz (\ansatz) for the field
requires
that we solve the boundary value problem,
\eqnam{\bvproblem}
$$\eqalign{&{d^2\phi \over d^2r} +{1\over r}{d\phi\over dr}-
{1\over r^2}\phi -\phi(\phi^2-1)=0\,,\cr
&\phi(r\to0)=0\,,\quad \phi(r\to\infty)=1\,.
}\eqno(\new)$$
This was solved numerically using relaxation techniques. The finite
boundary condition $\phi(r\to R) = 1$ introduces an error of order $R^{-2}$,
which we minimize using a large grid and spatial step-size. We can perturb
the straight string ansatz using the solutions defined in \S2 with relative
amplitude
$\varepsilon$. We  only impose perturbations in the $x$- and $y$-directions,
thus
introducing higher order errors, ${\cal O}(\varepsilon^2)$. However, these are
negligible for small amplitude perturbations and so we effectively create
highly accurate numerical solutions of the Nambu action.

It is appropriate at this point to comment on the
string self-field  and the
radiation field since they are not easily distinguished in the Goldstone
formalism.  There are, however,  particular configurations in which the two
can be
separated.  For example, when a perturbed string straightens the massless
self-field becomes azimuthal around the string as in (\ansatz) and the massive
self-field can be easily calculated numerically from (\bvproblem). In our
simulations,
this  allows a straightforward subtraction at specified times to yield both
the massless
and massive radiation fields.  Again, this procedure will be most accurate
for small
$\varepsilon$, though it still yields useful information for
$\varepsilon \sim 1$.

Naturally the spatial discretization introduces boundaries into  an
otherwise unbounded problem. These artifical boundaries have an important
effect on the
numerical solution, especially at late times in problems involving radiation.
For the
long string configurations on the $z$-boundary we impose simple periodic,
Dirichlet or
Neumann boundary conditions. As in ref.\refto{DSa}, on the $x$-
and $y$-boundaries we
employ absorbing boundary conditions, except here using a higher-order second
approximation version\refto{EM}\refto{EH}\refto{BJR} (refer to ref.
\refto{BSa}). These absorbing boundary
conditions require the solution of \eqnam{\boundarycond}
$${\partial\over\partial
t}{\partial\Phi\over\partial x} - {\partial^2\Phi\over \partial^2t} + {1\over
2}\left({\partial^2\Phi\over\partial^2y} + {\partial^2\Phi
\over\partial^2z}\right)=0\,,\eqno(\new)$$ at $x=0$ and $x=N_1\Delta x$
and a similar
equation at $y=0$ and $y=N_2\Delta x$. These remarkable boundary
conditions reflect a
wave incident at $45^{\circ}$ with only $3\%$ of the original amplitude,
that is, the
reflected wave energy is reduced by a factor of $10^{-3}$.
Fig.~6 illustrates the
efficacy of these boundary conditions by contrasting the
absence of reflected radiation
from absorbing boundary conditions with large reflections from  Neumann
boundary
conditions for the same $t$.  These allow the simulations to
be evolved for many grid
light-crossing times without significant influence on the interior string
solution.
Note that the grid face boundary conditions (\boundarycond) must be
appropriately
augmented along the grid edges and at the corners\refto{BSa}.

\doublefigure{neumann_inv1.eps}{absorb_inv1.eps}{1.7in}{-0.15in}{2.5in}{6}
{Comparison of (a) Neumann and (b) second approximation
boundary conditions for typical string radiation at late times}

Several runs were performed with Neumann boundary conditions on the $x$- and
$y$-boundaries to demonstrate energy conservation. Energy was conserved to
within
$0.2\%$ for both the 2-2 and 2-4 algorithms after several thousand timesteps.
Unfortunately, only first-order spatial differences were used in
calculating the energy
density so no marked diiference was evident between the two algorithms.
We should point out that the boundary conditions (\boundarycond) only
absorb massless
radiation, but strings emit both massless and massive radiation.
The massive radiation
in $\phi$ is not easily excited and it is highly suppressed at small amplitude.
Furthermore, because of dispersion, it does not reach  the boundary
until late times.
For reasonable values of $\varepsilon$ this massive radiation is
barely detectable and
has no discernible effect on string motion.

We usually employed a grid with $N_1=N_2=200$, $N_3=50$, and
spatial and time step-sizes
$\Delta x = 0.7$ and $\Delta t = 0.3$  (well within the
Courant condition for
these algorithms). Initial string configurations were
demonstrated to be numerically
stable for over 3000 timesteps, both in the interior and at the boundaries.
The
algorithm was used to study periodic helix, sine and kink long string
solutions and
circular loop collapse.  Throughout the simulations we monitored the
amplitude of the
perturbed oscillation $\varepsilon$ and the total energy per unit length $E$.
We also analyzed the massless radiation spectrum using the
three-dimensional spectral analysis techniques described in Appendix
D.

\figure{totalvrad.eps}{4in}{-0.15in}{7}{Spectrum for (a) the total
massless field
(b) the radiation field}

Our numerical results will be
seen to be at variance with those of Sikivie {\it et al.\/}
\refto{HarS,HagS,Sikb} for
three primary reasons: \vskip 0pt
\noindent (i) Firstly, in ref.\refto{HagS} they analyzed the total
massless field
spectrum, including the string self-field, which introduces non-radiating
higher
harmonics into their spectral analysis.  Fig.~7 demonstrates this clear
difference
between the total massless field spectrum and the  radiation spectrum
analysed after
subtration of the string self-field. Notice that the self-field
completely dominates  the radiation field at this amplitude.  Moreover,
the fall-off in
the higher modes for the total massless field spectrum is power-law
($P_n \propto
n^{-1}$), whereas for the radiation spectrum it is much more rapid,
being consistent with
exponential.
\vskip0pt \noindent (ii) Secondly, in ref.\refto{HagS} they
inadvertently concentrate on
degenerate configurations with $\varepsilon>1$ which do not
correspond to the Nambu
solutions presented in \S2.   These configurations are highly
degenerate and collapse
ultra-relativistically;  whole regions along the string would
achieve luminal
velocities\footnote{*}{These luminal string regions are
known as `lumps', and are
distinct from `cusps' where only  a single string point reaches
the speed of light.
Unlike ``lumps', cusps have a finite integrated radiation power
\refto{VVb}.\smallskip}
but for radiation damping.  An appropriate analogy is the
perfectly circular loop
solution which has a similar degeneracy and a corresponding
pathological radiation
divergence.   Additional difficulties with these configurations
are the inaccuracy of
the ansatz (\ansatz) at large amplitude and the fact that our
comparative analytic
results only apply in the $\varepsilon < 1$ regime.
As it happens for small simulation
scales, radiation damping for these ``luminal'' $\varepsilon > 1$
configurations
is so effective that higher harmonics are only marginally more excited.
Nevertheless, the $\varepsilon > 1$ results give the impression that the string
is critically- or over-damped, whereas the $\varepsilon < 1$ configurations
actually oscillate---as expected---like decaying Nambu solutions.
In any case, such degenerate configurations would
not be expected in a realistic string network, if only because of the
presence of small-scale structure, and they are not, therefore, a
reliable indicator of how it will radiate.
\vskip0pt
\noindent (iii) Thirdly, the reader is invited to compare the rather different
spectral analysis methods of ref.\refto{HagS} with those detailed in Appendix
D. In particular, notice that the standard definition of the one
sided three dimensional power spectrum (D3\warning) and the related wave number
(D2) are not
used in ref.\refto{HagS}. This apparent bin mislabelling will have an
important effect on
the numerically estimated power spectrum.

\figure{sine_evol.eps}{1.5in}{0.5in}{8}{Isosurfaces of the energy
density for the
evolving sine solution $(\varepsilon_0=0.6)$ shown after each period.
Radiation damping causes a noticeable
decrease in the oscillation amplitude.}

\subhead{Massless radiation from the sine solution}
\smallskip

The evolution of the sine perturbation (\cosinedef) was simulated for a number
of different initial amplitudes, ($\varepsilon_0=0.2,0.3,..,1.2$)
in order to obtain
a quantative understanding of the evolution. As expected we find that the
string
oscillates losing amplitude by emitting radiation.
Fig.~8 shows the decay of the maximum amplitude $\varepsilon$. Emitted
radiation is illustrated in fig.~9 in a plane transverse to the string at
the point of maximum amplitude.  The radiation pattern is dominated by
quadrupole lobes of the $n=2$ harmonic.

\doublefigure{sine_pict1_inv1.eps}{sine_pict2_inv1.eps}{1.8in}
{-0.15in}{2.5in}{9}
{Radiation quadrupole patterns in the plane perpendicular to the string
for $\varepsilon_0=0.6$
after (a) $t=3\lambda$ and (b) $t=7\lambda$. The difference the the
radiation field
form zero determines the colour saturation.}

\figure{sine_spectra_semi_same.eps}{3in}{-0.25in}{10}{Radiation power
spectrum for
the sinusoidal solution on a log-linear scale for (a) $\varepsilon_0=0.4$,
(b) $\varepsilon_0=0.6$ and (c) $\varepsilon_0=0.9$.}

We have analyzed the power spectra for $\varepsilon_0=0.4,\,0.6,\,0.9$;
these are
illustrated in fig.~10.  The subtraction of the string self-field from the
much smaller radiation field has to be performed precisely.  Nevertheless,
we found that some minor residuals could not be eliminated, notably in the
third and
fourth harmonics, because of the numerical inaccuracies inherent in this
procedure and
the nature of the radiation pattern near the string.\footnote{*}
{These appear to be
residuals because they remain when the oscillation energy and the
$n$$=$$2$ radiation
pattern have largely died away.  We believe that they are due to an imperfect
subtraction because the strings is not exactly straight at the
discretized sampling
timestep.  The effect is also exacerbated by some distortion due to
Lorentz contraction
and the fact that the radiation pattern is only purely quadrupole
asymptotically, not
in the immediate vicinity of the oscillating string.\smallskip}
Configurations producing the strongest radiation provide the cleanest spectra
($\varepsilon_0 =0.9$ in fig.~10c). We can only be confident that these results
provide an upper bound for the radiation in higher harmonics ($n>1$), but we
can
still draw definite conclusions.  First, it is clear that the lowest radiating
harmonic ($n$$=$$2$) provides the dominant radiation component, as predicted
analytically.  Secondly, although some higher harmonics appear to be present,
they
are strongly suppressed by radiation into the lowest harmonic---even for
$\varepsilon_0 \gapp 0.9$ they are not significantly excited.  This is an
important
observation which confirms the assumption underlying our simple
backreaction model
in section 2.    Finally, the decay  in the higher harmonics is
consistent, within
the uncertainties, with an exponential fall-off.  Without doubt it is
completely
inconsistent with a flat power law spectrum, $P_n \propto n^{-1}$.  An $n^{-1}$
fall-off would predict an amplitude for the $n$$=$$10$ harmonic relative to
$n$$=$$2$
which is nine orders of magnitude larger than that observed in fig.~10c.
The time evolution of the contribution to the radiation in the second harmonic
is
illustrated in fig.~11.  The energy increases until the radiation
propagates out to the edge of the box and begins to fall as the absorbing
boundaries take effect.

\figure{second_harm_small.eps}{3in}{-0.25in}{11}{The contribution of the
second fundamental mode
to the radiation spectrum for $\varepsilon_0=0.6$.}

\topinsert
{$$\hbox{\vbox{\offinterlineskip
\def\strut{\hbox{\vrule height 10pt depth 6pt width 0pt}}
\hrule
\halign{

\strut\vrule # \tabskip 0.1in &
\hfil#\hfil &
\vrule$\,$\vrule #&
\hfil#\hfil &
\vrule #&
\hfil#\hfil &
\vrule #&
\hfil#\hfil &
\vrule #\tabskip 0.0in \cr
& \hskip 0.2in $\varepsilon_0$ && $A$ && $B$ && $C$ &
\cr
\noalign{\hrule}\noalign{\hrule}
& \hskip 0.2in 0.3 && 28.18 &&  0.64 && 769.02 &
\cr\noalign{\hrule}
& \hskip 0.2in 0.4 && 28.14 &&  1.08 && 386.45 &
\cr\noalign{\hrule}
& \hskip 0.2in 0.5 && 28.08 &&  1.68 && 238.68 &
\cr\noalign{\hrule}
& \hskip 0.2in 0.6 && 28.02 &&  2.34 && 165.04 &
\cr\noalign{\hrule}
& \hskip 0.2in 0.7 && 27.98 &&  3.12 && 123.91 &
\cr\noalign{\hrule}
& \hskip 0.2in 0.8 && 27.90 &&  3.96 &&  99.37 &
\cr\noalign{\hrule}
& \hskip 0.2in 0.9 && 27.84 &&  4.88 &&  84.20 &
\cr\noalign{\hrule}
& \hskip 0.2in 1.0 && 27.78 &&  5.84 &&  74.82 &
\cr\noalign{\hrule}
& \hskip 0.2in 1.1 && 27.66 &&  8.00 &&  65.81 &
\cr\noalign{\hrule}
& \hskip 0.2in 1.2 && 27.50 && 10.42 &&  64.73 &
\cr\noalign{\hrule}}}}$$
\nobreak
\caption{Table 1}{Best fit values for the backreaction parameters
$A$, $B$ and $C$.}}
\endinsert

\mfigure{figure13.eps}{3.5in}{-0.25in}{12}{Energy against time:
Numerical results
and best fit curves for (a) $\varepsilon_0=0.5$, (b) $\varepsilon_0=0.8$ and
(c) $\varepsilon_0=1.1$.}

Evidence for quantitative agreement with the analytic radiation
calculations comes from
attempts to fit the time history of the overall energy density
to the simple backreaction
model (\brnonsoln). The total energy initially remains constant until the
radiation propagates
out to the boundary. At this stage it begins to fall with a
time-delayed profile similar
to that of our simple model.
The least squares estimators for $A$, $B$ and $C$ in (\brnonsolnen)
are calculated using a global minimisation routine and are
displayed in Table~1 for a
variety of initial amplitudes.

We find that $A$ is constant to within a few percent and the
calculated values for $B$ and $C$ are in good agreement for
$\varepsilon_0\lapp 1.1$. The numerical results and the best
fit curves are
shown in fig.~12 for $\varepsilon_0=0.5,\,0.8,\,1.1$.
A more significant test comes from determining the dependence
of $B$ and $C$ on
$\varepsilon_0$.  This is shown in  fig.~13 in a log-log plot, the points are
well-fitted by a straight line in the region
$0.3\lapp\varepsilon_0 \lapp 0.8$.  If we assume that
$A=A_0$, $B=B_0\varepsilon_0^{n_1}$ and $C=C_0\varepsilon_0^{n_2}$
then we calculate
least squares estimators, $A_0=28.0\pm 0.4$, $\log B_0=1.7\pm 0.2$,
$\log C_0=4.2
\pm 0.4$, $n_1=2.0\pm 0.1$ and $n_2=-2.0\pm 0.1$. These exponents
are in excellent
agreement with the predictions of the backreaction model (\brnonsolnen).
{}From these
values we can obtain the parameters in (\brenergy), $\alpha=0.2\pm 0.05$ and
$\beta=3.5\pm 1.75$. Given the uncertainties, these are again in satisfactory
agreement with the actual analytic values, $\alpha=0.25$ and $\beta=1.98$ from
(\cosinelinpower). We can also calculate the value of the unrenormalized string
tension by summing the massive field contributions to the energy of a straight
string solution, yielding $\mu_0=4.2\pm 0.2$.  This allows us to estimate
$\log(\Delta/\delta) =3.8\pm 0.2$.
This indicates that the simulated global strings have  approximately six
times more energy in the Goldstone field than in the massive string core. These
results, therefore, test and confirm the renormalization procedure which
underlies
the previous analytic predictions.

\figure{figure14.eps}{4in}{-0.15in}{13}{Loglog plots of (a) $B$ agianst
$\varepsilon_0$ and (b) $C$ against $\varepsilon_0$.}

We have also numerically studied the dynamics and radiation spectra of
sinusoidal solutions for $\varepsilon_0>1$ where the analytic results are
invalid.
These solutions are highly degenerate and were expected to produce a large
amount of radiation in higher radiation modes, however, this was not
found to be the
case. Fig.~14 shows  the radiation spectrum for $\varepsilon_0=2.0$,
showing that the
string predominantly radiates in the second harmonic even at these large
amplitudes. This result indicates that radiation backreaction damps the
excitation
of higher harmonics much more strongly than expected. On these small
simulation scales,
massive radiation is also excited at large amplitude (see below)
and this may contribute
to the radiation damping.

Finally, we should comment on the related helix solution which is predicted to
radiate extremely weakly (\helixlinpower).  Interestingly, we observe that the
helix appears to be dynamically unstable and radiates in a similar
manner to the sine solution (mainly in the $n=2$ harmonic). This may be
because our
initial ansatz is not a sufficiently perfect representation of the true
solution.
However, we believe that the non-radiating property of the helix is a
finely balanced
result of its symmetry.  Any asymmetry
introduced by radiation damping or otherwise will tend to
destabilize the solution
towards others which generically radiate more strongly. This
conjecture remains
to be justified analytically.

\figure{large_amp_spectra.eps}{4in}{-0.1in}{14}{The radiation spectrum for a
large
amplitude sine solution with $\varepsilon_0=2.0$. Notice the power in
the higher harmonics which shows up in the log-linear spectrum.}

\vfill\eject
\subhead{Massless radiation from the kink solution}
\smallskip

\figure{kink_evol.eps}{2in}{0.8in}{15}{Energy density isosurfaces  for
the kink solution $(\varepsilon_0=0.6)$ shown after each period of its
evolution.  Note the effectiveness of
`rounding' due to radiation damping.}

We have also studied the evolution of the kink solution (\kinkdef) for a
number of
different initial perturbation amplitudes. In section 2 we proposed
that the kink
solution should radiate with an $n^{-1}$ spectrum, truncated at
some time-dependent
`rounding' scale.  The effectiveness of this `rounding' can be
seen in fig.~15 which
shows an initial kink perturbation evolving into a sinusoidal perturbation.
The decay of high harmonics in the kink spectrum takes place
very rapidly. Fig.~16
shows the radiation spectrum at early times (a), contrasted
with the spectrum from late
times (b).  Note that the kink solution initially has more
power in  higher frequency radiation modes with a slow power
law fall-off.  The later
spectrum (fig.~16b), however, is indistinguishable from a typical sinusoidal
spectrum. These results certainly support our intuitive understanding of
kink `rounding'.

\mfigure{kinkvkink.eps}{3.3in}{-0.1in}{16}{The radiation spectra for the kink
solution after (a) $t=3\lambda$ and (b) $t=15\lambda$.
Compare the latter with the
sinusoidal spectrum given in fig.~7b.}

\subhead{Massive radiation}
\smallskip

We have already noted that global strings can
radiate massive particles, as well as massless particles. Depending on the
final
decay products, massive radiation from topological defects is of considerable
interest because it has been proposed as a potential source of baryon
asymmetry.
Massive modes will be suppressed except in regions of high curvature, such as
cusps and sharp kinks.  In our simulations because the curvature radius is
not much
larger than the string width, massive radiation could be excited for large
amplitude sinusoidal solutions, that is, $\varepsilon_0\gapp 0.6$.
Massive radiation produced by global strings exhibits very different
behaviour to
massless radiation.
Fig.~17 illustrates a circular pattern reminiscent of
dipole radiation, quite unlike the quadrupole patterns of fig.~9.
As the string
oscillates, massive radiation is emitted most strongly in the
direction of  string
motion. However, note that for amplitudes below
$\epsilon_0 < 0.9$ the massive radiation
signal was relatively very weak. Massive radiation will
be discussed in greater detail
elsewhere\refto{BSa}.

\doublefigure{massive_early_inv1.eps}{massive_late_inv1.eps}{2.0in}
{-0.25in}{2.5in}{17}{Massive radiation contours for a sine solution
with $\varepsilon_0=0.9$ after (a) $t=5\lambda$ and (b) $t=10\lambda$.
Note that
this is a weak signal with contours lying in the range, $-0.002 \lapp
1-\phi \lapp 0.002$.}

\figure{loop_evol.eps}{2.5in}{1in}{18}{Energy density isosurfaces
for a collapsing circular loop
shown at intervals of 20 timesteps.}

\mfigure{massless_loop_inv1.eps}{1.5in}{1.3in}{19}{Massless
radiation contours in
the aftermath of circular loop annihilation.  Illustrated
is a slice transvere to
the plane in which the loop collapsed; very little radiation
is emitted in the loop
plane.}

\subhead{Loop solutions}
\smallskip

The radiation spectrum from oscillating loops was not easily
accessible to numerical
simulation because on these small scales radiation damping
leads to their rapid demise.
There are further numerical difficulties setting up appropriate initial
configurations and distinguishing the radiation field from
the loop self-field.
Nevertheless, we illustrate the collapse of a perfectly
circular loop in fig.~18.
The loop begins to shrink rapidly under its own tension,
and annihilation occurs when
the opposite segments collide. At this stage massive
radiation is also produced
copiously.  The massless radiation pattern after
collapse is shown in fig.~19.  The
fact that the radiation is maximal in a direction
transverse to the loop plane is
consistent with analytic calculations for the
circular solution.  The predicted
divergence in the the integrated power is, of
course, removed by radiation backreaction
and finite width effects.  We note that this very
degenerate loop configuratoin is
expected to produce an atypical $n^{-1}$ power
spectrum, so we cannot infer general
conclusions for loop radiation from this special
case (as in ref.\refto{HagS}).

Our expectation on cosmological scales for
non-intersecting loop solutions is that
the antisymmetric tensor formalism of
ref.\refto{VVa} will provide an accurate
picture of loop radiation, that is generically
$P_n \propto n^{-\alpha}$ with
$\alpha \ge 3/2$ for $n>>1$.  In the final
stages of loop decay when its radius
approaches the core width, the demise will
proceed more rapidly through
annihilation as in fig.~18.  The resulting
small burst of massive radiation may
have significant cosmological implications.

\sectbegin{4~}{Discussion and conclusions}

The conclusions of this work are unambiguous.   We have made careful
comparisons between high precision numerical simulations of the
Goldstone model (\actionone) with analytic calculations using the antisymmetric
tensor formalism of the Kalb-Ramond action (\kraction).  The description
of vortex-line dynamics with both approaches is found to be in
close quantitative agreement.  This has a number of important implications,
not least because the validity and usefulness of this duality had previously
been questioned.

Firstly, this work confirms the procedure employed for passing between the
Goldstone boson field $\vartheta$ and the antisymmetric tensor field
$B^{\mu\nu}$ {\it in the presence of vortices}. Note that on the relatively
small scales we consider in these simulations, the duality is not
straightforward because the relation (\duality) is complicated by the
behaviour
of the massive field $\phi$ in the string core. Provided perturbation
wavelengths
are large compared to the string core width $\delta$, we can
integrate out the
massive degrees of freedom to obtain the Nambu term in
(\kraction). Of course, for
very high energy perturbations it proved possible to
excite these internal massive
degrees of freedom.

Secondly, the Nambu-like behaviour predicted for global
strings in (\renormeom) depends upon the renormalization of the string
self-field. (This is distinct from renormalization about a point source
because it involves cut-offs at both small and large scales.)
In our numerical
simulations, we typically had an energy in the string
self-field (the Goldstone mode)
which exceeded that in the core (the massive mode) by a
factor of about six.
Nevertheless, but for radiative damping, the evolution of
initially perturbed solutions
was observed to closely correspond to oscillatory Nambu
trajectories.  In no sense is
this self-field a diffuse `cloud of energy' somehow distinct
from the global string
core.  On the contrary, it is tightly bound to the
string---the nearer the core, the more
tightly.  On scales below the string curvature
radius, these massless modes
merely renormalize (and dominate) the string tension and energy density.

Thirdly, as predicted by the Kalb-Ramond action, global strings
radiate with a classical spectrum in harmonics directly related to the
wavelengths of perturbations on the string.  For example, for the sine
solution we confirmed the absence of the fundamental mode, demonstrating
that radiation was primarily in the second harmonic.
This discrete spectrum appeared to
be consistent with  exponential fall-off at high harmonics for the long
string configurations we studied, in complete contrast to the
prediction of power law
fall-off, $P_n \propto n^{-1}$, suggested in \refto{HarS,HagS,Sikb}.
Sharp kink
configurations initially showed additional power for large $n$,
but this was rapidly
eliminated by rounding due to radiation damping, and the spectrum soon
conformed to that for simple sine solutions.

Finally, we have numerically confirmed the quantitative rate of radiative
backreaction predicted for particular long string configurations.  Both the
amplitude and decay rate of the oscillations on the strings were found to be
in good agreement with our analytic results. Moreover, we observed
that radiation damping
was very effective at eliminating higher harmonics $n>2$, markedly
increasing the
amplitude range over which our simple backreaction model gave
satisfactory results.  We
are currently applying this work to more general string
configurations because it has
important implications for the understanding of string network evolution and,
specifically, the build-up and damping of small-scale structure on strings.
Modifications to (\renormeom) and direct comparisons with
field theory simulations are
the subject of a forthcoming publication\refto{BSb}.

It is interesting to place these results in a cosmological context because
this formed our initial motivation. For heavy cosmological
strings, the logarithm in
$\mu(\Delta)$ is much larger ($\ln (\Delta/\delta)\sim 70$--100)
than for strings in
our simulations.  Although the emitted power remains the same,
the overall decay rate
is slower because the radiation damping term in (\renormeom)
is relatively weaker.
Conversely, however, these strings can be expected to produce
more radiation in higher
harmonics because the string trajectory will more nearly
approach the strongly radiating
configurations allowed by the Nambu equations, notably cusps.

We believe that these results leave little doubt that the
low-energy effective Kalb-Ramond action provides an excellent
quantitative description
of global string dynamics and radiation.  Of course, as we
have observed  with massive
excitations, the antisymmetric tensor description needs to
be supplemented with the full non-linear field theory in
very high curvature regions,
such as at points of string intersection or cusps.
However, these regions typically have
length-scales corresponding to the string core width $\delta$,
so in a macroscopic
context they will have a very minor effect on overall predictions of massless
radiation.  In cosmology these results imply that the
antisymmetric tensor formalism should provide an accurate
picture of axion production by
an axion string network as described in ref.\refto{VVa,DSa,Sakb},
but contrary to the
assertions of refs.\refto{HarS,HagS,Sikb}.  Whether or not
this has been achieved with
sufficient precision at this stage remains an open question
which we briefly take up in
Appendix A.   However, axion strings are by no means the
only application of this
elegant and tractable formalism; global strings also
appear in low-temperature and other
contexts. Indeed, refs.\refto{DSc,GKPB} demonstrated the
close relationship between
global strings in the Goldstone model (\actionone) and
vortex-lines in a
`relativistic' superfluid. In ref.\refto{Car} this has been generalized to
compressible superfluids and perfect fluids, so  it may be that the methods
described here---both analytic and numerical---can be usefully applied
and tested
in a laboratory setting.

\nosectbegin{Acknowledgements}

\noindent We have benefitted from useful discussions with Mark Hindmarsh.
We are also
grateful for conversations with Brad Baxter, Brandon Carter and Mary
Sakellariodou.  We both acknowledge the support of the Science
and Engineering Research
Council, including the Cambridge Relativity group rolling grant (GR/H71550) and
Computational Science Initiative grants (GR/H67652 \& GR/H57585).

\nosectbegin{Appendix A: Axion string bounds revisited}

\eqnumber=1

The astrophysical and cosmological constraints on the axion were introduced in
section 1.  Here we consider the axion string bound in the light of our
previous
discussions, specifically noting the relative axion contributions radiated by
small loops and long strings.  Of course, we assume the standard axion scenario
with the universe passing through the Peccei-Quinn phase transition after
reheating.

Cosmic string evolution in an expanding universe has been extensively studied
numerically\refto{AT,BB,AShe}.  A long string network loses energy through loop
production and is observed to evolve toward a `scaling' regime in
which its large-scale
properties remain constant relative to the horizon.
These robust large-scale results were obtained for local gauge strings
but they can also be expected to apply to global strings.  The overall
long-string
energy density, however, must be modified with a logarithmic time dependence
arising through the cutoff ($\Delta\sim t$) in the global string energy density
$\mu$, that is,  \eqnam{\strgdensity}
$$
\rho_\infty = {\zeta\mu\over t^2} \approx {2\pi \zeta \fa^2\over t^2} \ln
(t/\delta)\,,\eqno(A\new)
$$
where  $\zeta \approx 14$ \refto{BB,AShe}. The network correlation length is
then given approximately by $\xi = \zeta^{-1/2}t$.

In numerical simulations, evolution
on small scales does not actually `scale', though this appears to have
little effect on
the overall density (A\strgdensity) and correlation  length $\xi$.
In any case, in
a realistic setting, this small-scale structure will be limited by
radiation backreaction, damping out perturbations below a constant
relative scale, $L_r
\equiv \kappa t$.  The long-string backreaction scale $L_r$ should
provide an lower
limit to the typical `scaling' size of loops created by the network, $\ell =
\alpha t$.

Axions will be radiated by loops and long strings until the
axion mass `switches on' at the time $t_{\rm w} = t_{\rm QCD}$ when
domain walls form, causing the rapid demise of the network.  We can
provide a greatly simplified estimate of the resulting number density of
axions $n_a$ by
noting that the dominant contribution comes from the
least relativistic axions
created just before $t_{\rm QCD}$\refto{Davb} (refer to ref.\refto{VS}).  To
maintain scaling in each Hubble time, the long-string network must
lose most of its
energy to loops  which, in turn, rapidly decay into axions
of roughly the same
wavelength.  Just prior to $t_{\rm w}$, then, loops will have made the
following
contribution to the axion energy density,  \eqnam{\loopaxdens}
$$
\rho_{{\rm a}\,\ell} \approx {2\pi \zeta \fa^2\over t_{\rm w}^2} \ln (t_{\rm
w}/\delta)\,.\eqno(A\new)
$$
These axions will typically have a frequency, $\omega =  4\pi/\alpha t$,
because loops of length $\ell\approx \alpha t$ radiate primarily in their
fundamental
mode $2/\ell$. For massless axions we have $\rho_{\rm a} =
\na \omega$, so the number
density of axions remaining at $t_{\rm w}$ is approximately
\eqnam{\loopaxnum}
$$
n_{{\rm a}\,\ell}\approx {\alpha \zeta \fa^2 \over 2t_{\rm w}}\ln (t_{\rm
w}/\delta) \,.\eqno(A\new)
$$
Clearly, a determination of the loop-size parameter $\alpha$ is essential
for a quantitative estimate of the axion density today.

Loop creation will occur on scales limited by the long-string radiation
backreaction scale, that is, $\alpha \gapp \kappa$.  Recalling the damping
`half-life'
$t_{1/2} = \alpha' \mu L/\beta \varepsilon_0^2$ defined in section 2,
we can determine
the minimum length-scale $L \approx \kappa t$ which will
have lost most of its
energy in one Hubble time:
\eqnam{\backreact}
$$
\kappa \sim {\pi^2\over 8} \left [ \ln (t/\delta) \right]^{-1}\,, \eqno(A\new)
$$
where we have taken $\beta = \pi^3 \fa^2/16$ and $\alpha'=0.25$
for the sine solution
(\sinedef), while also assuming $\varepsilon_0\approx 1$.
This yields $\kappa
\sim 0.02$ at $t_{\rm QCD}$, but large uncertainties must
be admitted. It seems reasonable, therefore, along with
previous authors to take
$\alpha\sim 0.1$\refto{VVa,VS}.

By equal matter--radiation $t_{\rm eq}$, the now-massive axions in
(A\loopaxnum) are
non-relat\-ivistic, so the axion density is given simply by
$\rho_{\rm a} \approx
\ma \na$, where $\na$ has been appropriately diluted by
the expansion $a^{-3}$.  A
comparison at $t_{\rm eq}$ between $\rho_{\rm a}$ and the critical density
$\rho_{\rm c}$ yields the cosmological
constraints\refto{VS},  \eqnam{\loopcon} $$
\ma \gapp 10^{-4}\,{\rm eV}\,,\qquad\fa \lapp 10^{11}\, {\rm
GeV}\,,\eqno(A\new)
$$
where we have assumed $\alpha \sim 0.1$.

A rough estimate of the axion contribution from long string radiation is also
straightforward.  According to (\cosinelinpower), the power radiated per unit
length is given by
\eqnam{\lstrgrad}
$$
{dP \over dz} = {\pi^3\fa^2 \over 16\gamma t}\,,\eqno(A\new)
$$
where we assume perturbations are peaked at some length-scale $\gamma t$
($\kappa <\gamma<\zeta^{-1/2}$), for which the relative amplitude
$\varepsilon \approx
1$. We now multiply by $t$ and the total length of string $\zeta/t^2$ to
obtain the axion energy density radiated in one Hubble time  prior to
$t_{\rm w}$,
\eqnam{\lstrgaxdens}
$$
\rho_{{\rm a}\, \infty} \approx {\pi^3\zeta \fa^2\over 16\gamma t_{\rm
w}^2}\,.\eqno(A\new)
$$
Given radiation in the second harmonic $\omega = 4\pi/\gamma t$,
we obtain the long
string axion number density
\eqnam{\lstrgaxnum}
$$
n_{{\rm a}\, \infty} \approx {\pi^2\zeta \fa^2\over 64 t_{\rm
w}}\,.\eqno(A\new)
$$
Note that the axion (or graviton etc.) number radiated by
long strings appears to be
independent of the actual perturbation length-scale which
dominates the spectrum.

A comparison between contributions from loops (A\loopaxnum) and long strings
(A\lstrgaxnum), yields the relative density ratio
\eqnam{\looplstrg}
$$
{\Omega_{{\rm a}\,\ell}\over \Omega_{{\rm a}\, \infty}} \approx
{32\alpha\over \pi^2}
\ln(t_{\rm w}/\delta)\approx {4\alpha\over \kappa}\,.\eqno(A\new)
$$
If loops are created above the backreaction scale $\alpha
\gapp \kappa$, then the axion density produced by long strings is probably
subdominant.  However, if $\alpha\sim \kappa$ these densities may be
comparable.
As mentioned in section 1, there are also other sources such as axion emission
during domain wall decay.  Again, approximate estimates suggest that this
contribution cannot be neglected in determining a final cosmological
bound on the
axion\refto{Lyta}.

\figure{axion_bounds.eps1}{3.45in}{0.2in}{20}{Schematic of the
approximate astrophysical
and cosmological bounds on the axion.  The viability of the
axion in the standard scenario clearly depends on an improved understanding of
axion string radiation. The conclusions of this paper indicate
that the true constraint
will lie somewhere near the string loop bound (\loopcon), leaving a a
somewhat smaller---but still substantial---parameter window.}

It remains to comment on the axion string constraints presented
in earlier work.
It is apparent that the bound (\loopcon) is weaker than the
original string bound
(\stringbound), $\fa\lapp 10^{10}{\rm GeV}$, of refs.\refto{Davb,DSa}.  The
difference lies not in the understanding of global string
radiation---which agrees
with this analysis---but, rather, in whether the dominant
network energy loss comes
from small loop creation or long string radiation.
The assumption\refto{Dava} that
it comes from long strings radiating at the correlation length $\xi \approx t$
seems to give an order-of-magnitude overestimate (comparable to taking the loop
result (A\loopaxnum) with $\alpha\approx 1$).
The apparent severity of this old
long string constraint is illustrated in fig.~20.

The difference between (A\loopcon) and the string bounds of
refs.\refto{HarS,HagS,Sikb} are more fundamental because
they are based on the
assumption that global strings radiate with a flat power spectrum,
$P_n \propto
n^{-1}$.  This is contrary to our results which show that power is generally
dominated by the lowest allowed harmonic.  The remaining literature on
cosmological axion constraints has been flawed either
by prevarication on this
issue or by ignoring topological effects altogether.
Authors have preferred to
reserve judgment on the underlying global string spectrum, while discussing the
consequences of both alternatives (see, for example, refs.\refto{Lyta,KT}).

This paper only constitutes a first step towards a definitive axion string
constraint.  While it supports the assumptions on which (A\loopcon) is based,
we
make no suggestion that axion detection experiments should either be launched
or
cancelled on the basis of the uncertain parameter window shown in fig.~20.
However, we do hope to remove these remaining uncertainties with large-scale
string network simulations which incorporate radiation backreaction.

\nosectbegin{Appendix B: Power expression split with left- and right-movers}

\eqnumber=1

One can split arbitrary solutions of the Nambu equations into
left- and right-moving
waves by employing null coordinates on the string worldsheet
$u=\sigma -\tau$ and
$v=\sigma -\tau$ (ie. $X^{\mu}=X_R^{\mu}+X_L^{\mu})$.
This was first used
by Burden\refto{Bur} to separate the Fourier transform
of the energy--momentum tensor
into left- and right-moving components when calculating
the gravitational radiation power
from a closed loop (see a more extensive discussion in
\refto{AShe2}). This split
has also been performed for a long periodic string
\refto{Hina}; unfortunately,
however, there was a slight oversight which has important
consequences for the
allowed spectrum of radiation.

Here we outline the left- and right-splitting of the Fourier transform of
the source distribution $\widetilde J^{\mu\nu}$ for
a $z$-directed long global string, periodic in $\sigma$ and
$\tau$ with period $L$ and
$X^3=\alpha\sigma$ (conditions satisfied by the helix solution
(\helixdef)). We then
discuss the generalization to the case where $X^3=\alpha\sigma
+ P_1(u)+ P_2(v)$ such
that $P_1$ and $P_2$ are arbitrary (small) functions with period $L$.
In the special
case of $X^3=\alpha\sigma$, we have \eqnam{\sourcedist}
$$\eqalign{J^{\mu\nu}({\bf x}^{\perp},z,t)=&{f_a\over 2}
\int_{-\infty}^{\infty}d\sigma
\int_{-\infty}^{\infty}d\tau (\dot X^{\mu} X^{\prime\nu}
- X^{\prime\mu}\dot X^{\nu})
\cr&~~~\times\delta^2({\bf x}^{\perp}-{\bf
X})\delta(z-\alpha\sigma)\delta(t-\tau)\,,\cr \widetilde J^{\mu\nu}({\bf
k}^{\perp},\kappa_m,\omega_n)=&{1\over \alpha L^2} \int_0^Ldt\int_0^{\alpha
L}dz\int\nolimits d^2{\bf x}^{\perp}J^{\mu\nu}({\bf x}^{\perp} ,z,t)
e^{-ik.X}\,,}
\eqno(\hbox{B}\new)$$
for $n\ne 0$ and $k^{\mu}=({\bf k}^{\perp},\kappa_m,\omega_n)$
where $\omega_n={2\pi
n/ L}$ , $\kappa_m={2\pi m/\alpha L}$. Using these definitions
one can deduce that
\eqnam{\sourceft}
$$\eqalign{\widetilde J^{\mu\nu}({\bf k}^{\perp},\kappa_m,
\omega_n)&={f_a\over
2\alpha l^2}
\int_0^Ld\sigma\int_0^Ld\tau(\dot X^	{\mu} X^{\prime\nu}
- X^{\prime\mu}\dot X^{\nu})
e^{-ikX}\,,\cr
&\equiv\int_0^Ld\sigma\int_0^Ld\tau\,I\,.}\eqno(\hbox{B}\new)$$
We now change the variables of integration from $(\sigma,\tau)$ to the null
coordinates $(u,v)$, in which $I$ can be re-expressed  as
\eqnam{\chnagevar}
$$\dot X^{\mu} X^{\prime\nu} - X^{\prime\mu}\dot X^{\nu} = 2\bigg{(}
\partial_vX_L^{\nu}\partial_uX_R^{\mu}-\partial_uX_R^{\mu}
\partial_vX_L^{\nu}\bigg{)}\,.\eqno(\hbox{B}\new)$$
The region of integration is more complicated. Fig.~21 shows how the
integration over
the region $0<\sigma <L$ and $0<\tau <L$ (that is, $R_1\cup
R_2\cup R_3\cup R_4$)
can be transformed into an integration over the region
$0<u<L\hbox{ and }0<v<2L$
($R_1\cup R_2^{\prime}\cup R_3^{\prime}\cup R_4$). This can be further
simplified by  noting that
\eqnam{\region}
$$ \int_{R_4\cup R_2^{\prime}}dudv\,I = e^{i(m+n)\pi}
\int_{R_1\cup R_3^{\prime}}dudv\,I
\eqno(\hbox{B}\new)$$
Therefore we can write the Fourier transform of the source
distribution function in terms
of left- and right-movers as
\eqnam{\leftright}
$$\widetilde J^{\mu\nu}({\bf k}^{\perp},\kappa_m,\omega_n)
={f_a\over \alpha}\big{[}
U^{\mu}V^{\nu}-V^{\mu}U^{\nu}\big{]}\eqno(\hbox{B}\new)$$
where
\eqnam{\defufull2}
$$\eqalign{U^{\mu} &= \int_0^L{du \over L} \partial_u
X_R^{\mu}e^{-ik.X_R(u)}\,,\cr
V^{\mu} &= \int_0^L{dv \over L} \partial_v
X_L^{\mu}e^{-ik.X_L(v)}\,,\cr
k^{\mu} &= (\omega_n,{\bf
k}^{\perp},\kappa_m)\,.}\eqno(\hbox{B}\new)$$
where $m+n$ must be even. This condition was missed in
the previous derivation and has a
important effect on the allowed spectrum of
radiation. In particular it is responsible
for  there being no contribution from the
fundamental harmonic ($n$$=$$1$).

\figure{left_right.eps1}{2.5in}{0.25in}{21}{
Equivalent regions of integration for $(\sigma
,\tau)$ and $(u,v)$.}

The generalisation of this splitting to a solution with $X^3=\alpha\sigma
+P_1(u)+P_2(v)$  can be achieved, assuming the $P_1\hbox{ and }P_2$
are sufficently small
by allowing the  lines bounding the regions $R_1,R_2,R_3$ and
$R_4$ to become
oscillatory. The  periodicity of fig.~21 is maintained by the periodicity of
$P_1$ and $P_2$.

\nosectbegin{Appendix C: Formulae for gravitational radiation power}

\eqnumber=1

This appendix gives gravitational radiation power expressions
for local strings,
analogous to the those for Goldstone bosons from global strings. The source
term  for the gravitational radiation is the energy--momentum tensor
\eqnam{\energymom}
$$T^{\mu\nu} = \mu_0\int (\dot X^{\mu}\dot X^{\nu} -
X^{\prime\mu}X^{\prime\nu})
\delta^{(4)}[x-X(\sigma,\tau)]\,
d\sigma d\tau\,,\eqno(\hbox{C}\new)$$
and the radiation power is given by
\eqnam{\gravrad}
$$\eqalign{{dP \over dz} = 2G\sum_{n=1}
^{\infty}\omega_n\sum_{m\,st\, |\kappa_m|<\alpha\omega_n}
\int_0^{2\pi}d\varphi\bigg{(}&\widetilde{T}^{\mu\nu*}
(\omega_n,{\bf k}^{\perp},\kappa_m)
\, \widetilde{T}_{\mu\nu}(\omega_n,
{\bf k}^{\perp},\kappa_m)\,\cr
&-{1 \over 2}|T^{\lambda}_{\lambda}(\omega_n,{\bf k}^{\perp},\kappa_m)|^2
\bigg{)}\,.}\eqno(\hbox{C}\new)$$
On performing the split into left- and right-movers, as in appendix B,
one can deduce
that
\eqnam{\gravradsplit}
$${dP \over dz} = {32\pi G\mu_0^2 \over \alpha^2L}
\sum_{n=1}^{\infty}n\sum_{|m|<
\alpha n \atop  m+n\,{\rm even}} \int_0^{2\pi}
d\varphi \big{\{}|U|^2|V|^2 +
|U^*.V|^2 - |U^.V|^2\big{\}}
\,,\eqno(\hbox{C}\new)$$
and the corresponding linearized power expression is given by
\eqnam{\gravradlin}
$${dP \over dz} = {64\pi G\mu_0^2 \over L}\sum_{n=1}^
{\infty}n\sum_{|m|<
 n \atop  m+n\, {\rm even}} \big{\{}|{\bf U}^{\perp}|^2|
{\bf V}^{\perp}|^2 -
|{\bf U}^{\perp *}.{\bf V}^{\perp}|^2 + |{\bf U}^{\perp}.{\bf V}^{\perp}|^2
\big{\}}\,.\eqno(\hbox{C}\new)$$
Note the difference between the previous expression (C\gravradsplit) and the
linearized version (C\gravradlin) (a correction for ref.\refto{Hina}).
Applying
(C\gravradlin) to the helix, sine, kink and sine spectrum solutions
discussed in section 2 yields the following
\eqnam{\gravpowersoln}
$$\eqalignno{
\hbox to 1in{Helix:\hfil} &{dP\over dz} = {4\pi^2G\mu_0^2\varepsilon^4 \over
L}\,,&(\hbox{C}\new)\cr
\hbox to 1in{Sine:\hfil} &{dP\over dz} = {\pi^2G\mu_0^2\varepsilon^4\over
2L}\,,&(\hbox{C}\new)\cr
\hbox to 1in{Kink:\hfil} &{dP\over dz} = {65536G\mu_0^2\varepsilon^4 \over
\pi^6L}\sum_{n=1}^{\infty}n \sum_{|m|<n \atop m+n\,even}{1 \over
(n^2-m^2)^2}\,,&\cr &~~~~~~~~\times\bigg{[}1 - \cos {\pi (m+n) \over
2}\bigg{]}\bigg{[}1 - \cos{\pi (m-n) \over 2}\bigg{]} \,,&(\hbox{C}\new)\cr
\hbox to 1in{Sine spectrum:\hfil}&{dP\over dz} = {4\pi^2 G\mu_0^2
\varepsilon^4\over
L}\sum_{n=1}^{\infty}\sum_{|m|<n  \atop m+n\,even}\bigg{[}|{\bf A}_{m+n \over
2}|^2|{\bf B}_{m-n \over 2}|^2&\cr  &~~~~~~~~~~~~ +|{\bf A}_
{m+n \over 2}.{\bf B}_{m-n
\over 2}|^2 - |{\bf A}_{m+n \over 2}^*.{\bf B}_{m-n \over 2}|^2\bigg{]}
\,.&(\hbox{C}\new)\cr}$$  This behaviour is very similar to
antisymmetric tensor
radiation. The notable exception is  the helix solution
which radiates gravitationally
in the second harmonic, underlining the incongruous
nature of the Goldstone boson result.
Note also that the linearized power expressions above
are only valid for $\varepsilon< \varepsilon_{\rm crit}$
for some $\varepsilon_{\rm
crit}= {\cal O}(1)$.  A simple backreaction model for
gravitational radiation will
take the same form as (\brnonsolnen).

\nosectbegin{Appendix D: Three-dimensional spectral analysis}

\eqnumber=1

\noindent Standard methods for power spectrum analysis
\refto{W,H,M} are applied to
the radiation field on a three-dimensional grid.
This involves calculation of the
three-dimensional discrete Fourier transform using an FFT algorithm.
The three-\-dimen\-sional Fourier transform, $\widetilde F_
{\beta_1,\beta_2,\beta_3}$,
of some discrete function $F_{\alpha_1,\alpha_2,\alpha_3}$ is given by
\eqnam{\ftdef}
$$ \widetilde F_{\beta_1,\beta_2,\beta_3} = {1 \over \sqrt{N_1N_2N_3}}
\sum_{\alpha_1=1}^{N_1-1}\sum_{\alpha_2=1}^{N_2-1}\sum_{\alpha_3=1}^{N_3-1}
F_{\alpha_1,\alpha_2,\alpha_3}e^{-2i\pi \big{(}{\alpha_1\beta_1\over N_1}
+{\alpha_2\beta_2\over N_2}+{\alpha_3\beta_3\over N_3}\big{)}}\,,
\eqno(\hbox{D}\new)$$
\noindent where
\eqnam{\fapprox}
$$\eqalign{&F_{\alpha_1,\alpha_2,\alpha_3} \simeq
F(\alpha_1\Delta x,\alpha_2\Delta y,
\alpha_3\Delta z)\,,\cr
&\widetilde F_{\beta_1,\beta_2,\beta_3} \simeq \widetilde
F\bigg{(}{2\pi\beta_1\over N_1\Delta x},
{2\pi\beta_2\over N_2\Delta x},{2\pi\beta_2\over N_3\Delta x}\bigg{)}\,,\cr
&0\leq\alpha_i,\beta_i\leq N_i-1\hbox{ for i=1,2,3}\,.}\eqno(\hbox{D}\new)$$
Given the above, the one-sided power spectrum is defined as\refto{PFTV}
\eqnam{\powerdef}
$$ P_{\alpha_1,\alpha_2\alpha_3} = {1 \over (N_1N_2N_3)^2}\bigg{[}
|\widetilde F_{\alpha_1,\alpha_2,\alpha_3}|^2+|\widetilde
F_{N_1-\alpha_1,N_2-\alpha_2
,N_3-\alpha_3}|^2\bigg{]}\,,\eqno(\hbox{D}\new)$$
\noindent for $1\leq\alpha_i\leq({N_i\over 2}-1)$ i=1,2,3.
In the special case when
one or more of the $\alpha_i$ is either zero or ${N_i \over 2}$,
then adjustments
must be made; in particular,
\eqnam{\powerdefone}
$$\eqalign{&P_{0,0,0} = {2 \over (N_1N_2N_3)^2}|\widetilde F_{0,0,0}|^2\,,\cr
&P_{{N_1 \over 2},{N_2 \over 2},{N_3 \over 2}} = {2 \over (N_1N_2N_3)^2}
|\widetilde F_{{N_1 \over 2},{N_2 \over 2},{N_3 \over 2}}|^2\,,}
\eqno(\hbox{D}\new)$$
\noindent and similar expressions are used for $P_{0,\alpha_2,\alpha_3}\,,\,
P_{0,0,\alpha_3}$ etc.  Once we have the three-dimensional
one-sided power spectrum, we
calculate the modes to which the power spectrum corresponds. This is done by
comparing  the wavelength of the string with the wave number
of each point in the power
spectrum and putting the contribution into an appropriate bin.

There is a problem associated with this approach to power spectrum analysis.
The
discrete Fourier transform assumes that the initial data is
periodic, which is not
necessarily the case for the radiation field in our simulations. The resulting
discontinuities at the boundaries can cause spurious harmonics to appear
in the spectrum.
We overcome this by multiplying the data by
a window function, $W(x)$. Window functions should be periodic and
fall-off quickly
toward zero  at the edges. To apply the window function we
calculate the discrete
Fourier transform of $G(x) = W(x)F(x)$. Using the convolution
theorem and the properties
of $W(x)$, we can recover the Fourier transform of $F(x)$ as
\eqnam{\window} $$ \widetilde G_{\beta_1,\beta_2,\beta_3} =
{\widetilde
F_{\beta_1,\beta_2,\beta_3} \over W_{ss}} \,,\eqno(\hbox{D}\new)$$
\noindent where
\eqnam{\sumsq}
$$W_{ss} = \sum_{\alpha_1=1}^{N_1-1}\sum_{\alpha_2=1}^{N_2-1}
\sum_{\alpha_3=1}^{N_3-1} \bigg{(}{W_{\alpha_1,\alpha_2,\alpha_3} \over
N_1N_2N_3}\bigg{)}^2 \,.\eqno(\hbox{D}\new)$$
The window function $W_{\alpha_1,\alpha_2,\alpha_3}$ is decoupled in each
direction, that is, $W_{\alpha_1,\alpha_2,\alpha_3}=
W_{\alpha_1}W_{\alpha_2}W_{\alpha_3}$, where each of the $W_{\alpha_i}$ is a
one-dimensional window function similar to those described in ref.\refto{H}.
After
extensive trials, it appeared that the problem was best
suited to  a Gaussian window function, that is,
\eqnam{\gaussian}
$$W_{\alpha_i}=e^{-\beta\left(\alpha_i-{N_i\over 2}\right)^2}\,,
\eqno(\hbox{D}\new)$$ where
$i=1,2,3$ and $\beta$ is comparable to the inverse of the
simulation box-length.


\def\hang{}


\def\jnl#1#2#3#4#5#6{\hang{#1 [#2], `#3,' {\it #4\/} {\bf #5}, #6.}
									}



\def\jnltwo#1#2#3#4#5#6#7#8#9{\hang{#1 [#2], `#3,' {\it #4\/} {\bf #5}, #6;
{\it #7\/} {\bf #8}, #9.}
									}

\def\prep#1#2#3#4{\hang{#1 [#2], `#3,' #4.}
									}

\def\proc#1#2#3#4#5#6{\hang{#1 [#2], `#3', in {\it #4\/}, #5, eds.\ (#6).}
}
\def\procu#1#2#3#4#5#6{\hang{#1 [#2], `#3', in {\it #4\/}, #5, ed.\ (#6).}
}

\def\book#1#2#3#4{\hang{#1 [#2], {\it #3\/} (#4).}
									}

\def\genref#1#2#3{\hang{#1 [#2], #3}
									}


\def\prl{Phys.\ Rev.\ Lett.}
\def\pr{Phys.\ Rev.}
\def\pl{Phys.\ Lett.}
\def\np{Nucl.\ Phys.}

\def\apj{Ap.\ J.}

\def\cup{Cambridge University Press}

\nosectbegin{References}

\references

\refis{Lyth1}
\prep{Lyth, D.H.}{1993}{Dilution of cosmological densities by saxino decay}
{Lancaster preprint}

\refis{SN}
\jnl{Frieman, J., Dimopoulous, S. \& Turner, M.S.}{1987}{Axions and Stars}
{\pr}{D36}{2201}

\refis{RG}
\jnl{Dearbon, D.S.P., Schramm, D.N., Steigman, G.}{1986}
{Astrophysical constraints on
the coupling of axions, majorons and familons}{\prl}{56}{26}

\refis{CKLL}
\jnl{Copeland, E.J., Kolb, E.W., Liddle, A.R., \& Lidesey, J.E.}{1993}
{Observing
the inflaton potential}{\prl}{71}{219}

\refis{BSa}
\prep{Battye, R.A., \& Shellard, E.P.S.}{1993}{Numerical field
theory simulation methods
for global topological defects}{in preparation}

\refis{BSb}
\prep{Battye, R.A., \& Shellard, E.P.S.}{1993}{String radiation backreaction}
{in preparation}

\refis{LR}
\jnl{Lund, F., \& Regge, T.}{1976}{Unified approach to strings \& vortices
with soliton solutions}{\pr}{D14}{1525}

\refis{Tur}
\jnl{Turok, N.}{1984}{Grand unified strings and galaxy
formation}{\np}{B242}{520}

\refis{Bur}
\jnl{Burden, C.J.}{1985}{Gravitational radiation from a particular class
of cosmic
strings}{\pl}{164B}{277}

\refis{EM}
\jnl{Engquist, B., \& Majda, A.}{1977}{Absorbing boundary conditions for
the numerical
simulation of waves}{Mathematics of Computation}{31}{629}

\refis{EH}
\jnl{Engquist, B., \& Halpren, L.}{1990}{Long time behaviour of absorbing
boundary
conditions}{Mathematical Methods in the Applied Sciences}{13}{189}

\refis{BJR}
\jnl{Bamberger, A., Joly, P., \& Roberts, J.E.}{1990}{Second order
boundary conditions for the
wave equation : A solution for the corner problem}{SIAM Journal of
Numerical Analysis}{27}
{323}

\refis{W}
\jnl{Welch, P.D.}{1967}{The use of fast fourier transform for the
estimation of power spectra :
A method based on time averaging over shorth, modifed periodograms}
{IEEE Transactions on Audio
and Electroacoustics}{AU-15}{70}

\refis{H}
\jnl{Harris, F.L.}{1978}{On the use of window for harmonic analysis
with the discrete
fourier transform}{Proceedings of the IEEE}{66}{51}

\refis{M}
\jnl{McClellan J.H.}{1982}{Multidimensional spectral estimation}
{Proceedings of the IEEE}
{70}{1029}

\refis{PFTV}
\book{Press, W.H., Flannery, B.P., Teukolsky, S.A., \& Vetterling,
W.T.}{1986}{Numerical recipes}{\cup}

\refis{Car}
\prep{Carter, B.}{1993}{Kalb-Ramond coupled string
vortex dynamics in a relativistic
superfluid}{DARC preprint}

\refis{AT}
\jnl{Albrecht, A., \& Turok, N.}{1989}{Evolution of cosmic string
networks}{\pr}{D40}{973}

\refis{BB}
\jnl{Bennett, D.P., \& Bouchet, F.R.}{1990}{High resolution
simulations of cosmic
string evolution: network evolution}{\pr}{D41}{2408}

\refis{AShe}
\jnl{Allen, B., \& Shellard, E.P.S.}{1990}{Cosmic string evolution---a
numerical simulation}{\prl}{64}{119}

\refis{AShe2}
\jnl{Allen, B., \& Shellard, E.P.S.}{1992}{Gravitational radiation from a
cosmic string network}{\pr}{D45}{1898}

\refis{AS}
\jnl{Abbott, L.F., \&  Sikivie, P.}{1983}{A cosmological bound on the
invisible axion}{\pl}{120B}{133}

\refis{PQ}
\jnltwo{Peccei, R.D., \& Quinn, H.R.}{1977}{CP conservation and the
presence of pseudoparticles}{\prl}{38}{1440}{\pr}{D16}{1791}

\refis{HagS}
\jnl{Hagmann, C., \& Sikivie, P.}{1991}{Computer simulations of the motion and
decay of global strings}{\np}{B363}{247}

\refis{HarS}
\jnl{Harari, D., \& Sikivie, P.}{1987}{On the evolution of global strings
in the early universe}{\pl}{195B}{361}

\refis{Sika}
\jnl{Sikivie, P.}{1982}{Axions, domain walls and the early universe}
{\prl}{48}{1156}

\refis{Sikb}
\genref{Sikivie, P.}{1992}{`Dark matter axions', Proceedings of First I.F.T.\
Workshop on {\it Dark Matter}, University of Florida, Gainseville, Feb.\
14--16,
1992.}

\refis{Dava}
\jnl{Davis, R.L.}{1985}{Goldstone bosons in string models of galaxy
formation}{\pr}{D32}{3172}

\refis{Davb}
\jnl{Davis, R.L.}{1986}{Cosmic axions from cosmic strings}{\pl}{180B}{225}

\refis{DSa}
\jnl{Davis, R.L., \& Shellard, E.P.S.}{1989}{Do axions need
inflation?}{\np}{B324}{167}

\refis{DSb}
\jnl{Davis, R.L., \& Shellard, E.P.S.}{1988}{Antisymmetric tensors and
spontaneous symmetry breaking}{\pl}{214B}{219}

\refis{DSc}
\jnl{Davis, R.L., \& Shellard, E.P.S.}{1989}{Global strings and superfluid
vortices}{\prl}{63}{2021}

\refis{Sa}
\procu{Shellard, E.P.S.}{1986}{Axionic domain walls \& cosmology}
{{\rm Proceedings
of the 26th Liege International Astrophysical Colloquium,}
The Origin and Early
History of the Universe}{Demaret, J.}{University de Liege}

\refis{Sb}
\proc{Shellard, E.P.S.}{1990}{Axion strings and domain walls}{Formation and
Evolution of Cosmic Strings}{Gibbons, G.W., Hawking, S.W., \& Vachaspati,
V.}{\cup}

\refis{Sc}
\jnl{Shellard, E.P.S.}{1987}{Cosmic string interactions}{\np}{B283}{624}

\refis{Sd}
\procu{Shellard, E.P.S.}{1992}{The numerical study of topological defects}
{Approaches
to Numerical Relativity}{d'Inverno, R.}{\cup}

\refis{PWW}
\jnl{Preskill, J., Wise, M.B., \& Wilczek, F.}{1983}{Cosmology of the invisible
axion}{\pl}{120B}{127}

\refis{Wil}
\jnl{Wilczek, F.}{1978}{Problem of strong $P$ and $T$ invariance in
 the presence of instantons}{\prl}{40}{279}

\refis{Wei}
\jnl{Weinberg, S.}{1978}{A new light boson?}{\prl}{40}{223}

\refis{KR}
\jnl{Kalb, M.  \& Ramond, P.}{1974}{Classical direct interstring
action}{\pr}{D9}{2273}

\refis{KT}
\book{Kolb, E.W., \& Turner, M.S.}{1990}{The Early Universe}{Addison-Wesley,
Redwood City, California}

\refis{VE}
\jnl{Vilenkin, A., \& Everett, A.E.}{1982}{Cosmic strings and domain walls in
models with Goldstone and pseudo-Goldstone bosons}{\prl}{48}{1867}

\refis{VVa}
\jnl{Vilenkin, A., \& Vachaspati, T.}{1987}{Radiation of Goldstone bosons from
cosmic strings}{\pr}{D35}{1138}

\refis{VVb}
\jnl{Vachaspati, T., \& Vilenkin, A.}{1985}{Gravitational radiation from cosmic
strings}{\pr}{D31}{3052}

\refis{Wit}
\jnl{Witten, E.}{1985}{Cosmic superstrings}{\pl}{153B}{243}

\refis{Hina}
\jnl{Hindmarsh, M.B.}{1990}{Gravitational radiation from kinky
infinite strings}{\pl}{251B}{28}

\refis{Hinb}
\genref{Hindmarsh, M.B.}{1993}{private communication}

\refis{Saka}
\jnl{Sakellariadou, M.}{1990}{Gravitational waves emitted from infinite
strings}{\pr}{D42}{354}

\refis{Sakb}
\jnl{Sakellariadou, M.}{1991}{Radiation of Nambu--Goldstone bosons from
infinitely long cosmic strings}{\pr}{D44}{3767}

\refis{VS}
\book{Vilenkin, A. \& Shellard, E.P.S.}{1993}{Cosmic strings and other
topological defects}{\cup ({\it in press})}

\refis{PRS}
\jnl{Press, W.H., Ryden, B.S., \& Spergel, D.N.}{1989}{Dynamical
evolution of domain walls in an expanding universe}{\apj}{347}{590}

\refis{DQ}
\jnl{Dabholkar, A., \& Quashnock, J.M.}{1990}{Pinning down the
axion}{\np}{B333}{815}

\refis{GKPB}
\jnl{Gradwohl, B.,  Kalbermann, G., Piran, T., \& Bertschinger,
E.}{1990}{Global strings and superfluid vortices: analogies and
differences}{\np}{B338}{371}

\refis{Lyta}
\jnl{Lyth, D.H.}{1992}{Estimates of the cosmological axion density}{\pl}
{275B}{279}

\endreferences

\vfill
\end